\documentclass[manuscript,nonacm]{acmart}

\usepackage{hyperref}

\AtBeginDocument{%
  \providecommand\BibTeX{{%
    \normalfont B\kern-0.5em{\scshape i\kern-0.25em b}\kern-0.8em\TeX}}}

\setcopyright{acmcopyright}
\copyrightyear{2023}
\acmYear{2023}

\usepackage{subcaption}

\begin{document}

\title{Preemption Aware Task Scheduling for Priority and Deadline Constrained DNN Inference Task Offloading in Homogeneous Mobile-Edge Networks}

\author{Jamie Cotter}
\email{jamie.cotter@mycit.ie}
\affiliation{%
    \department{Computer Science}
  \institution{Munster Technological University}
  \streetaddress{Rossa Avenue}
  \city{Cork}
  \country{Ireland}
  \postcode{T12 P928}
  }

\author{Ignacio Casti\~{n}eiras}
\email{ignacio.castineiras@mtu.ie}
\affiliation{%
    \department{Computer Science}
  \institution{Munster Technological University}
  \streetaddress{Rossa Avenue}
  \city{Cork}
  \country{Ireland}
  \postcode{T12 P928}
}

\author{Donna O'Shea}
\email{donna.oshea@mtu.ie}
\affiliation{%
    \department{Computer Science}
  \institution{Munster Technological University}
  \streetaddress{Rossa Avenue}
  \city{Cork}
  \country{Ireland}
  \postcode{T12 P928}
}

\author{Victor Cionca}
\email{victor.cionca@mtu.ie}
\affiliation{%
    \department{Computer Science}
  \institution{Munster Technological University}
  \streetaddress{Rossa Avenue}
  \city{Cork}
  \country{Ireland}
  \postcode{T12 P928}
}

\begin{abstract}
Deep Neural Networks (DNNs) are highly computational applications frequently applied to use-cases such as photogrammetry, video and image processing and often possess strict latency requirements. Recent work has explored offloading DNN tasks to nearby devices with similar processing capabilities, to avoid the larger communication delays incurred for cloud offloading.

We consider a three-stage image/video processing pipeline with 1) object detection, 2) low complexity classifier, 3) high complexity classifier, where 1) and 2) are always run locally and 3) can be offloaded, and 3) has lower priority and can be preempted. 
Devices will prioritize the execution of their own tasks. By accepting workloads from other devices, there is a risk that the local tasks will not receive the resources required to complete on time.
We present a preemption aware scheduling approach for priority and deadline constrained task offloading in homogeneous edge networks. Our scheduling approach consists of two distinct scheduling algorithms, designed to accommodate the differing requirements of high and low priority tasks. To satisfy a task's deadline, our scheduling approach considers the availability of both communication and computational resources in the network when making placements in both the current time-slot and future time-slots. The scheduler implements a deadline-aware preemption mechanism to guarantee resource access to high priority tasks. When low-priority tasks are selected for preemption, the scheduler will attempt to reallocate them if possible before their deadline. Task resource requirements are derived from offline and online measurements of communication and processing times. 

We implement this scheduling approach into a task offloading system which we evaluate empirically in the real-world on a network of edge devices composed of four Raspberry Pi 2 Model B's. We evaluate this system under against a version without a task preemption mechanism as well as workstealing approaches to compare the impact on high priority task completion and the ability to complete overall frames. These solutions are evaluated under a workload of 1296 frames.

Our findings show that our scheduling approach allows for 99\% of high-priority tasks to complete while also providing a 3 - 8\% increase in the number of frames fully classified end-to-end over both workstealing approaches and systems without a preemption mechanism.
\end{abstract}



\received{N/A}
\received[revised]{N/A}
\received[accepted]{N/A}

\maketitle

\section{Introduction}

Deep Neural Networks (DNNs) are highly computationally expensive applications used to solve challenges with problem spaces too large to model using traditional approaches. DNN inference tasks are frequently applied to use-cases such as photogrammetry \cite{effectdnn}, video \cite{adamask} and image processing \cite{edgeoffloadimagerate} and often possess strict latency requirements. IoT and edge oriented deep learning frameworks such as Tensorflow Lite have made it feasible to process DNN tasks locally. Use-cases such as image classification often have a high rate of task generation which can quickly overwhelm an edge devices capacity to process them, especially if several tasks are generated simultaneously.
However, if nearby edge devices are idle, offloading can help alleviate local capacity constraints \cite{review_comp_offloading}. Task offloading requires that some or all of the workload is sent from one device to another. If each device in the network is executing the same type of tasks then only the input to the application will be transferred, otherwise the task binary will also be transferred. The main issue in computational offloading is allocating machines to process tasks and scheduling when they should execute them, which may be further complicated by constraints such as deadlines or locality.

Image classification applications at the edge are often structured as a set of tasks composing a pipeline. Similar to other works \cite{classificationpipeline}, this paper considers a three stage pipeline consisting of: 1) an object detector 2) low complexity classifier and 3) a stage comprised of one or more highly computational classification tasks that are generated in parallel. Each stage in the pipeline can potentially classify the frame, terminating processing early or otherwise progressing further into the pipeline until each stage is exhausted \cite{distream, polly}. When stage three tasks are offloaded to another edge device, it is important that stage two tasks on the selected edge device are given priority access to the resources they need as they must be processed locally due to deadline constraints.

To mitigate remote stage 3 tasks blocking resource access to local higher priority tasks, we use preemption to eject the blocking task allowing high priority tasks immediate access to the resources they need. 

In this paper we present a solution for deadline-constrained task offloading in mobile-edge networks by developing a preemption aware scheduling algorithm that considers following: the workload capacity of devices, the communication on the network link, the priority of tasks, and finally, the deadline of tasks offloaded. Our scheduling algorithm uses a preemption mechanism to guarantee that high priority tasks receive the resources they require in times of resource scarcity.

We have evaluated this approach in real-world experiments on Raspberry Pi 2 Model Bs \footnote{Raspberry Pi 2 Model B spec sheet: https://docs.rs-online.com/790d/0900766b8139232d.pdf}, as processors that execute a three-stage waste image classification pipeline. We have compared this system against a version of the system which does not contain a preemption mechanism as well as a workstealer based solution. Based on the results of our experiments, our findings are the following:
\begin{itemize}
    \item Preemption leads to an overall increase in processed frames end-to-end, from a 3\% $-$ 8\% improvement.
    \item The cost of preemption leads to 3\% $-$ 20\% less DNN tasks completing in each late stage pipeline from .
    \item Preemption allows $10\% - 23\%$ more high-priority tasks to complete vs non-preemption capable approaches, resulting in a 99\% completion rate of high priority early stage tasks.
    \item Preemption increases the number of tasks in the network, which leads to a slight degradation in the time to allocate early and late stage tasks in scheduled solutions.
    \item Schedulers outperform workstealers in processing constrained pipeline applications under preemption conditions, completing approximately 23\% more frames than workstealers.
\end{itemize}

The paper continues with related work in both deadline oriented offloading and DNN partitioning in Section \ref{sec:background}. Then, Section \ref{sec:horizontal} presents the design of the system and horizontal partitioning, the benefits it brings and the problem it introduces. Section \ref{sec:algorithms} discusses the two primary algorithms implemented in our system. Then, Section \ref{sec:implementation} presents the experiment implementation. The results of the experiments are evaluated in Section \ref{sec:evaluation}, followed by a discussion of the results in Section \ref{sec:discussion}. Finally, Section \ref{sec:conclusions} provides some conclusions.

\section{Related Work}
\label{sec:background}

Kang et al. \cite{neurosurgeon} propose a partitioning solution for DNN offloading in MEC environments. The motivation for their work came from the high communication overheads incurred by offloading DNN tasks to cloud. The communication latency could often be 11 times larger than
computational latency due to the size of the input data. Their solution was to introduce a "vertical" split into the model where it is partially processed locally to reduce the amount of data sent to the cloud and reduce communication overheads. They devised a lightweight scheduler to identify the optimal split in the DNN that minimised the end-to-end latency. a portion would be processed locally and the remainder in the cloud to reduce communication latency.

Zhao et al. \cite{deepthings} present a solution for DNN distributed DNN processing in resource-constrained edge networks. They utilise a task posting scheme. To mitigate the poor computational performance of these edge devices they implement a technique they call FTP (fused tile partitioning). FTP allows for parallel processing of a chain-based DNN while also implementing the vertical partitioning technique seen in the work of Kang et al.

Zeng et al. \cite{distream} propose a deadline aware distributed system for processing video analytic pipelines in smart-camera edge deployments. Their approach factors in the hardware heterogeneity of edge devices, potential workload imbalances that can occur between the cameras and the edge cluster and the variance in the workload in the network over time. In their proposed solution smart cameras can perform partial processing of the video pipeline locally or offload it to nearby smart cameras with the inference stage offloaded to the edge cluster. They utilise an LSTM to predict the incoming workloads across devices which is then fed into a stochastic partitioning scheme and finally they routinely adapt the workload across devices when an imbalance is detected. They identify the optimal partitioning scheme which is formulated as an optimisation problem with the objective to maximise system throughput with respect to pipeline deadlines.

Nigade et al. \cite{jellyfish} present a deadline constrained DNN offloading framework for edge networks. Their use case focuses on offloading from mobile-edge devices to an edge server where the inference tasks will be processed. They consider soft deadline constraints which clients provide to the system as service level objectives.
They utilise DNN adaption to satisfy throughput constraints where their scheduler periodically maps clients to a DNN model of a certain input size based on the expected network bandwidth and the processing time of the model with respect to the soft-deadline outlined in the SLO.

Zhou et al. \cite{adaptiveparallelexecution} present a framework for DNN partitioning and offloading in heterogeneous edge networks. Their framework provides a novel spatial partitioning technique that improves upon previous solutions by allowing for the full model to be partitioned rather than the first seven layers. Additionally their framework implements a dynamic programming solution to find the optimal partition and parallelisation strategy based upon the available computation and communication resources in the network.

Gupta et al. \cite{dnnschedulingedgecloud} propose a dual-objective heuristic for deadline-constrained heterogeneous DNN offloading applied to video segment processing generated by UAV drones in a VIP monitoring scenario. UAV drones transfer video segments to a resource constrained edge accelerator where multiple applications can choose to create inference processing requests on the video segment. Their scheduler maintains two task queues, an edge queue and a cloud queue. When scheduling a task, the scheduler checks to see if the task will meet its deadline on the edge and evaluates the impact to existing tasks in the queue, if a task has a higher priority than existing tasks then any tasks that would violate their deadline as a result are relegated to the cloud queue. Additionally, they implement a work stealing mechanism allowing the edge device to steal work from the cloud queue if there has been slack generated from edge tasks completing earlier than expected.

Bari et al. \cite{dynamicadaptivejobs} propose a deadline oriented offloading policy for MEC scenarios. They focus on a scenario where a set of users offload data for image detection DNN models to a base station for processing. They structure input data as a set of layers, where each layer is associated with an offloading cost based on the input size and a compatible model. Offloading further data layers in the set will allow for an increased detection quality at the cost of increased transfer time. Their goal is to maximise the number of layers offloaded across all users with respect to each job's deadline.

Zeng et al. \cite{coedge} present CoEdge a cooperative inference system for DNN offloading among resource-constrained heterogeneous edge devices. Their goal is to minimise system energy costs while also satisfying application latency requirements. Their system  considers hard deadline constraints and exploits model parallelism through the use of horizontal partitioning to partition a DNN in order to reduce the processing time of the application. They devise their partitioning scheme adaptively based on available devices, current network conditions and the expected energy cost.

Based on the existing literature, while there are solutions that tackle hard deadline constraints and priority constrained task processing in edge networks, there is an open gap in solutions that guarantee resource access to deadline-constrained high priority tasks in homogeneous mobile edge environments.
\section{System Design}
\label{sec:system_model}

\begin{figure}[h]
\centering
\begin{subfigure}{0.49\textwidth}
\includegraphics[width=\textwidth]{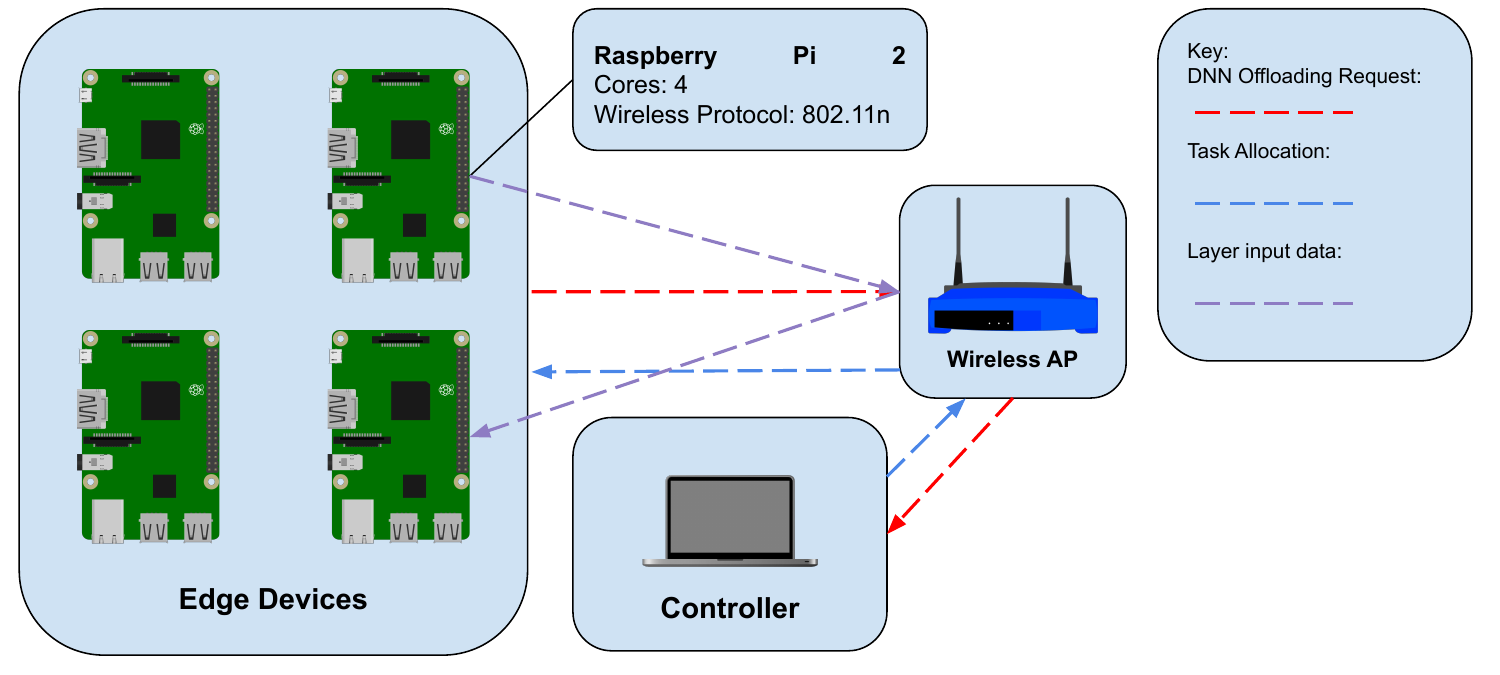}
\caption{Network Topology}
\label{fig:network_topology}
\end{subfigure}
\hfill
\begin{subfigure}{0.49\textwidth}
\centering
\includegraphics[width=\textwidth]{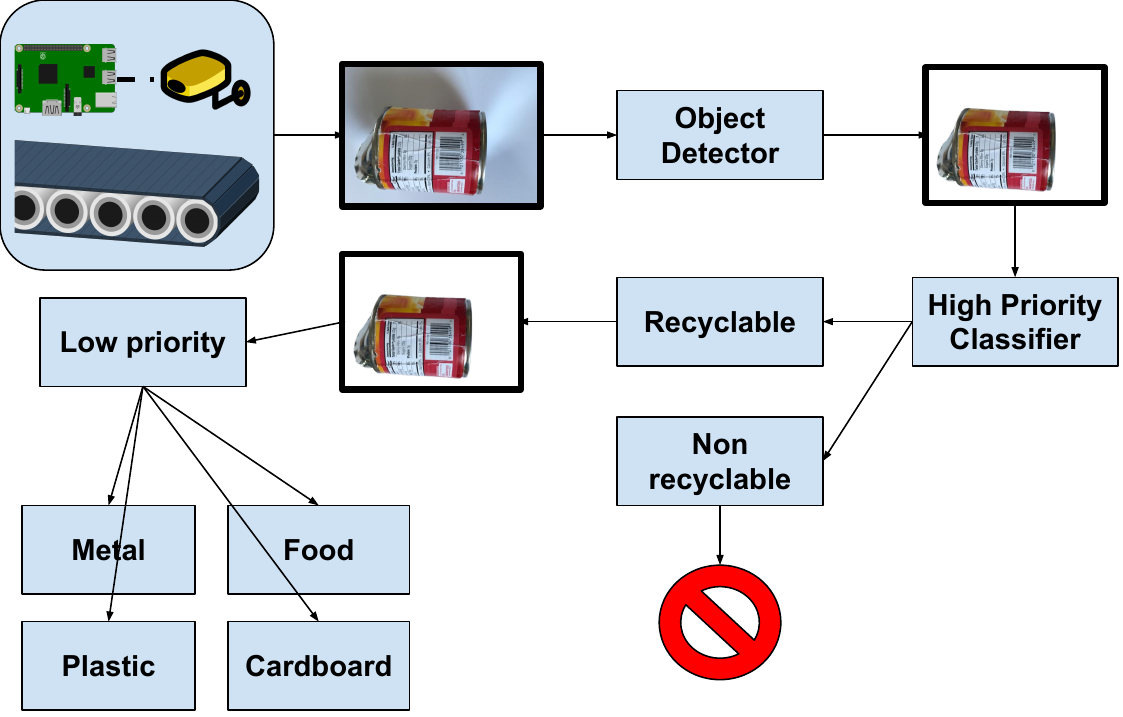}
\caption{Task pipeline}
\label{fig:pipeline}
\end{subfigure}
\caption{System design}
\end{figure}

This work considers the waste classification system in Fig. \ref{fig:network_topology} where a set of four devices monitor waste items on their conveyor belt running at constant speed with the goal of sorting waste into recyclable classes, sampling the pipeline at a fixed rate. The irregular distribution of waste on a conveyor belt can result in several classification tasks or even none if there are no recyclable objects detected in the frame.

The waste classification pipeline shown in Fig. \ref{fig:pipeline} has three stages that enable classification of refuse into recycling classes. The first step detects if an object is present on the conveyor belt. Considering a uniform colour conveyor belt, the object detection can be implemented as a simple foreground detection operation. Object detection runs for all frames. If an object is detected it will be classified into general waste or recyclable using a \textit{low-complexity} binary classifier. We trained a Support Vector Machine (SVM) on SIFT features of the TrashNet \cite{trashnet} data-set. The final stage is run only if the object is classified as recyclable, and uses a \textit{high-complexity} Convolutional Neural Network (CNN) to further classify into four classes of recyclable waste. For the CNN the YoloV2 network was used and trained using TrashNet.

The timings of the three stages were measured on the RPi2B at: 100ms for stage 1 and 980ms for stage 2. The final stage takes more than 20s on the RPi2B. This can be reduced by parallelising the processing over several cores using horizontal partitioning \cite{deepthings}, splitting the input to convolutional layers into smaller tiles. The timing of the final stage then depends on the number of cores over which it is parallelised and the existing workload on the machine at the time of processing.

We focus only on the two classification tasks because object detection can be considered a constant processing overhead. Considering the order of the stages, the stage 2 low-complexity classifier into recyclable or not is assigned \textit{high priority}, while the final high-complexity classification is considered \textit{low priority}. From an application point of view it is more important to complete the stage 2 than stage 3.

The workload across devices is not uniform in nature. Depending on the number and type of waste items present on the conveyor belt a device is assigned to, it is possible that a device may be processing nothing while others are incapable of processing the entirety of their own task load. Computational offloading can help mitigate the problem of imbalanced workloads and make better use of idling resources in the network \cite{distream}. The hosts are forced to start as pairs in a staggered fashion. This is to further increase the load imbalance and improve the ability to offload. The pairs examine frames in different time cycles. Two checking at the start of the cycle and the other two at middle cycle, this ensures that two hosts can potentially spawn high-priority tasks while the other two devices are generating low-priority tasks. However, edge devices are not synchronised and there will be a random offset between any two devices at the start of a frame.

The system considers a network structured as a star topology composed of the controller, computational hosts and a shared network link routed through the wireless AP. The controller is responsible for all decision making in the network. The controller uses a timeslotted allocation approach for communication and computational resources such that no two tasks are allocated the same resource simulataneously \cite{femtoclouds}. To account for tasks and messages of variable size, the time-slots are of variable length \cite{deadlienconstrainedoffloading}. The controller receives task allocation requests (both stage two and three) from edge devices and makes appropriate placement decisions based on the constraints of the task type and the availability of network resources. The controllers perception of network state is maintained by tracking placement decisions and the result of executed tasks.

All time-slots that are created by the scheduler include additional time-padding at the end of created time-slots to account for real-time performance variation. 
To obtain our time-slot padding we use the standard deviation of performance tests for processing padding and the jitter in the network tests as communication padding.

Additionally, the controller is responsible for 
measuring average network bandwidth to make accurate time-slotted reservations on the network link.

\subsection{Communication Interactions}
Edge devices communicate to the controller when they request task placement for stage two and stage three tasks that must be executed by a certain deadline as well as the result of executed tasks.
The controller communicates to edge devices to inform them of scheduling decisions, as well as preemption decisions. If a task is successfully allocated, the device is informed if the task will be processed locally or if the input must be transferred to another device for remote processing.

Inter-edge device communication occurs when an edge device is offloading a low-priority task to another device, where it must transfer the input image for the model to process.

High priority tasks are always executed locally on the device that generated the initial offloading request,only require one CPU core to process and are always allocated at the time they enter the scheduler at. Therefore, to allocate a high priority task into the network, it requires the following:
an initial time-slot on the link to inform the device whether or not its workload was allocated and if so, a time-slot to process on its allocated device and finally a time-slot on the link to send a status update to the controller.

Low priority tasks can use up to several CPU cores and have the option of being offloaded which impacts the set of time-slotted resources they will require. To allocate a low priority task to its local device they require the following:
an initial time-slot on the link to inform the device of its workload, a time-slot to process on its allocated device, a time-slot on the link to send a status update to the controller and finally the number of CPU cores the task requires.
Finally, if a low priority task is offloaded to another device, it requires an additional time-slot on the link to exchange its input image to the allocated host.

\subsection{Horizontal partitioning}
\label{sec:horizontal}
To reduce the processing time of the low priority Yolov2 classifier (final detection blocks of YoloV2 are not included) the inference is partitioned using horizontal partitioning and parallelised between several cores. The number of cores, and respectively the number of partitions of the input are decided by the scheduler. YoloV2 is a sequence of several blocks consisting of convolutional layers followed by a max-pooling layer. Based on \cite{hadidi2020toward} partitions of input data can be processed through several consecutive convolutional blocks by expanding the partitions around the edges to account for the expansion of the convolutional filter, and discounting the overlapping borders when combining the results. In some cases, the separate processing of input partitions can continue through max-pooling layers. However, if the expansion border associated with convolution does not align with the stride of the max-pooling layer this will lead to loss of data. For the generalised case max-pooling layers must therefore process the entire output of the previous convolutional block, rather than individual partitions. This means that each partition is processed through a consecutive block of convolutional layers, then the outputs are combined into an intermediate output which is processed by the following max-pooling layer, and the subsequent output is ready again for partitioning and parallel processing.

To avoid network communication delays a single DNN is horizontally partitioned and distributed over the different cores of a single device. Nevertheless, the partitioning and assembly of features between convolutional and max-pooling blocks creates Inter-Process Communication overheads that can be significant. We reduced these overheads to a minimum after observing that between convolutional and max-pooling blocks only the border of a tile changes, while the inner part stays the same, therefore only the border must be communicated.

In this system we utilise a two-core and a four-core partitioning scheme. When a single DNN task is generated at the low-priority stage it can be executed in the four-core configuration, receiving maximum reductions to its processing time. However, when several DNN tasks are generated simultaneously, the system must now decide whether or not to offload a task and whats its core configuration will be, based on the following.

\begin{itemize}
    \item offloading additional tasks leads to communication delays incurred but allows for tasks to be executed at higher core configurations resulting in lower processing time while also impacting remote devices.
    \item executing multiple tasks locally at a lower core configuration, incurring no communication overheads but longer processing times.
\end{itemize}

\subsection{Request Processing}
\label{sec:system_design}

As seen in Fig. \ref{fig:network_topology}, the system implements a master-worker architecture. Edge devices issue task requests to the controller which then allocates resources to process the task in the network.
Incoming task placement requests (both stage 2 and stage 3) are placed in an internal job queue upon arrival alongside system tasks related to maintaining the internal network state representation or scheduling sub-routines such as the preemption mechanism. Messages are processed by priority and arrival time within their priority class.
If a stage 2 task invokes preemption on a stage 3 task then it will return the stage 3 task to the job queue for re-processing.
Stage 2 tasks can be processed in parallel as they are constrained to their local device and a device can only generate a single stage 2 task at a time. Otherwise, all requests and jobs in the queue are processed in a blocking sequential fashion.

\section{Algorithms}
\label{sec:algorithms}
In this work we present a scheduling approach for DNN task offloading in homogeneous edge networks that aims to prioritise network resources to high priority tasks in times of resource contention.

To satisfy this problem, we must allocate each task time-slotted windows that satisfy their communication and computation requirements. Additionally, any allocated time-slots must complete before the tasks deadline, otherwise any result generated will bring no benefit to network. Finally, when higher priority tasks enter the network, they must be guaranteed access to these network resources even at the cost of lower priority tasks.

To satisfy the differing requirements of our two types of tasks with respect to their priority, we present two main subroutines:
\begin{itemize}
    \item High priority allocation algorithm.
        \begin{itemize}
            \item With preemption mechanism.
            \item Without preemption mechanism.
        \end{itemize}
    \item Low priority allocation algorithm.
\end{itemize}

The high priority algorithm first finds the earliest time-slot that can accommodate the allocation message on the network link with respect to available throughput and existing link allocations. Next, the scheduler calculates the processing time-slot [t1, t2] by using the time the allocated message is expected to arrive on the edge device as t1 and t2 = t1 + the benchmarked processing time. If the total core usage of existing tasks that overlap with the processing time-slot plus the additional core for the high priority task does not exceed the source devices capacity then the task is allocated. Otherwise the high-priority task is not allocated. 
If preemption is enabled and allocation is not possible the scheduler must generate a preemption request for the source device at this time-slot. 

Unlike high-priority tasks, low-priority tasks can be offloaded to other devices and executed in various configurations, which can reduce processing times but increase core usage. A successful high-priority allocation request always results in a single task being allocated, whereas low-priority requests may generate multiple tasks. For a low-priority request to be considered complete, all of these tasks must execute successfully within their request's deadline.

The low-priority scheduler operates over a set of time points, representing the completion of existing tasks and the release of their occupied resources back into the network. This set is constrained to time-points between the moment the scheduler is called until the request deadline.

At each time point, the scheduler attempts to allocate any remaining unallocated tasks from the initial request. The scheduler first reserves the network link for the allocation message as early as possible and allocates a time window for image transfer (in case the task is offloaded). Next, the scheduler searches for a device that can process a given task at the minimum viable resource configuration (e.g. two-cores) within the processing window, resulting in a partial allocation if successful.

When selecting a device for partial allocation, the scheduler prioritises the task's source device to avoid the need for image data transfer. If that is not possible, it aims to distribute tasks evenly across devices in the network. After attempting a partial allocation for each unallocated task, the scheduler then tries to improve each task's allocation by reducing processing time, checking if the allocated device can support increased resource usage.

Finally, for each allocated task, the scheduler reserves a state update message on the network link.

This process repeats until every task in the request has been allocated, or all time points have been exhausted.

When the high-priority scheduler fails to allocate a high-priority task, it begins the preemption process, where it iterates over the tasks source device and selects a single conflicting task with the farthest deadline for preemption. It then re-runs the high-priority scheduler for the failed task and finally attempts to reallocate the preempted low-priority task by searching for a device can execute it before its deadline.

\section{Implementation}
\label{sec:implementation}

The edge devices we have chosen for our experiments are Raspberry Pi 2 Model Bs. The controller was run on a 2020 Macbook Pro (M1).

The controller runs the scheduling algorithms which were implemented in C++ 20 (compiled using clang 16.0.6); the edge devices run inference managers implemented in Python 3.11.4, which are responsible for receiving allocated tasks from the controller and processing them or offloading them to another edge-device's inference manager. The DNN tasks are processed using TensorFlow lite \footnote{https://www.tensorflow.org/lite} and our DNN model is based upon YoloV2.

Estimates of network communication throughput and jitter are required to make accurate scheduling decisions on the network link for image transfer associated with task offloading. We obtain these estimates on system startup using Iperf3, each edge device hosts an Iperf server which are contacted by an Iperf client on the controller on system start-up. Due to offloading within the local network our throughput was effectively halved as all communication has to route through the AP. In our first experiment (with preemption) the average throughput was $\sim16.3$MBps and in the second experiment (no preemption) the average throughput was $\sim18.78$MBps.

To ensure time synchronisation between the controller and the four edge devices, we hosted an NTP server on the controller and configured each edge device to use the controller as their NTP server. Additionally, we used the library NTPlib in python to manually fetch a synchronised time-stamp from the server on system startup and manually set the system clock of the edge devices.

Communication between the edge-devices and the controller is implemented using REST. On the edge devices we implemented a basic REST interface using the Python HTTP package and on the controller we used cpprestsdk \footnote{https://github.com/microsoft/cpprestsdk}. Input images are transferred between edge devices using multipart uploads, we use multipart uploads so that if data transfer is interrupted we only need to upload the chunk that failed to transfer.

We have obtained estimated max-sizes for each type of communication message in the network via benchmarking (high priority task allocation: 700 bytes, low priority allocation: 2250 bytes, state update: 550 bytes, preemption message: 550 bytes and input transfer 21500 bytes). The actual duration of the time-slots for each of these communications is dependent on the initial throughput estimates generated at system start-up.

Each task and task configuration (ie. high-priority, low priority with two-cores and four-cores) have fixed processing times based upon our benchmark tests which result in fixed-length time-slots for each stage and configuration type. The values are as follows: high-priority (0.98s), two-core low-priority (16.862s) and four-core low-priority (11.611s).
To minimise the impact of system load and hardware variations during run-time on our low-priority high-complexity DNN tasks we use the standard deviation from benchmark tests as padding on the processing time.

For the experiments, each edge-device also runs a secondary Python application which is our experiment manager. It generates the frames for the pipeline at regular intervals (corresponding to the conveyor belt speed and determining the pipeline processing deadlines). If the controller can successfully allocate a high-priority task then its execution is simulated by having the experiment manager sleep for the allotted window.
If a high-priority task is determined to have spawned a set of low-priority tasks, it issues a low-priority request to the controller that can contain 1 to 4 DNN tasks. If the controller can allocate these tasks' resources, it then forwards their allocation onto the inference manager of the chosen hosts.

For the purposes of our experiment set-up we use the same input image for each DNN task. DNNs require inputs to be of a certain size. In real-world scenarios, waste items would be extracted from the source frame and resized before sending them to the DNN.

To evaluate the various levels of task load we model the experiment using trace files. These configurations control the distribution of high priority and low priority tasks (DNNs) generated per frame. Each entry in a trace file represents workload for four devices in a given frame. Where a device in a frame can have one of the following values: -1 (no object is detected), 0 (a high-priority task is generated but with no low-priority request afterward) and 1 .. 4 (a high-priority task generated and a low-priority request with n number of DNN tasks is generated after it completes). In our experiments we use five different trace files representing different distribution of generated DNN tasks: in uniform devices generate 1..4 tasks with equal probability; in weighted X (x in 1..4) devices will predominantly generate X tasks, with the network load increasing with X.

A new task pipeline is generated every ~18.86 seconds, we derived this value from running this system with the minimum viable completion time for the object detector, a high-priority task and one low-priority DNN task partitioned across two cores.

For comparison, we also implement two workstealing solutions, a decentralised workstealer in which each device maintains their own queue of generated low-priority tasks and must poll other edge devices for work and a centralised workstealer where edge devices generate low-priority tasks and post them to a centralised job queue on the controller which other edge devices can then steal from.

Finally, we compare multiple versions of the system. One with preemption enabled and one without against the centralised and decentralised workstealer with and without preemption. We evaluate the workstealers under the weighted four scenario specifically.

\section{Results}
\label{sec:evaluation}

To evaluate our system we focused on several key performance metrics, the main one being the frame completion rate as this is the overall indicator of whether or not preemption is a net positive. However, to understand fully as to why systems perform at the level they do we must analyse other factors related to the completion and resource usage of high and low priority requests and the role preemption played in their completion. For low priority tasks we are interested in the difference between raw completion rate and set completion rate, where the latter refers to all the tasks generated by the same high priority task.

We perform the evaluation under several workloads, comparing the performance of a scheduler against a centralised and decentralised workstealer approach which are known to provide near-optimal results in unconstrained work scheduling problems. We run these experiments on each solution with and without preemption to determine the impact that it has on the above metrics.

Finally, we also measure the run time of the preemption scheduler under different loads.

In doing so we hope to address the following: Preemption allows additional high priority tasks to enter the network, which in turn generate their own workloads; does this create a cascade effect where the resource scarcity leads to even more high priority tasks requiring preemption? Does the lack of coordination in workstealing approaches lead to a higher rate of preemption when compared to their scheduled counterparts? Finally, does the increased volume of allocated tasks in the network negatively impact the search time for the system scheduler?

\begin{table}[]
\begin{tabular}{l|l}
\textbf{Experiment}                                                                              & \textbf{Label} \\ \hline
Uniform Scheduler Preemption                                                                     & UPS            \\ \hline
Uniform Scheduler Non-Preemption                                                                 & UNPS           \\ \hline
\begin{tabular}[c]{@{}l@{}}Weighted N (1 .. 4) - Preemption \\ Scheduler\end{tabular}            & WPS\_N         \\ \hline
\begin{tabular}[c]{@{}l@{}}Weighted 4 - Non-Preemption \\ Scheduler\end{tabular}                 & WNPS\_4        \\ \hline
\begin{tabular}[c]{@{}l@{}}Weighted 4 - Decentralised \\ Preemption Workstealer\end{tabular}     & DPW            \\ \hline
\begin{tabular}[c]{@{}l@{}}Weighted 4 - Decentralised \\ Non-Preemption Workstealer\end{tabular} & DNPW           \\ \hline
\begin{tabular}[c]{@{}l@{}}Weighted 4 - Centralised \\ Preemption Workstealer\end{tabular}       & CPW            \\ \hline
\begin{tabular}[c]{@{}l@{}}Weighted 4 - Centralised \\ Non-Preemption Workstealer\end{tabular}   & CNPW          
\end{tabular}
\caption{Graph Legend}
\label{tab:graph_legend}
\end{table}

\subsection{Frame Completion}
\begin{figure}
    \centering
\begin{subfigure}{0.49\textwidth}
    \includegraphics[width=\textwidth]{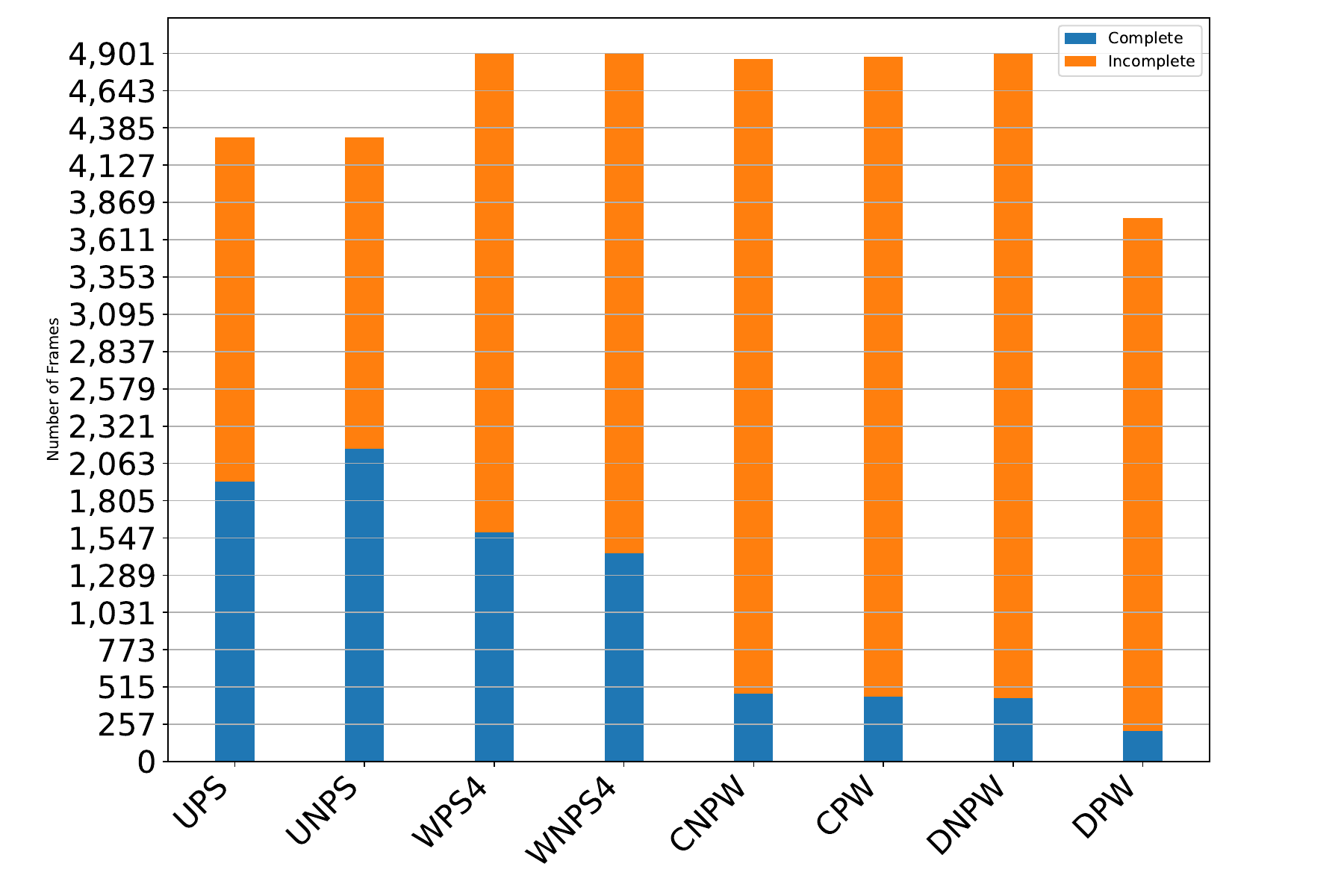}
    \caption{Preemption vs. non-preemption} 
    \label{fig:preempt_vs_non_preempt_frame_Completion}
\end{subfigure}
\begin{subfigure}{0.49\textwidth}
    \centering
    \includegraphics[width=\textwidth]{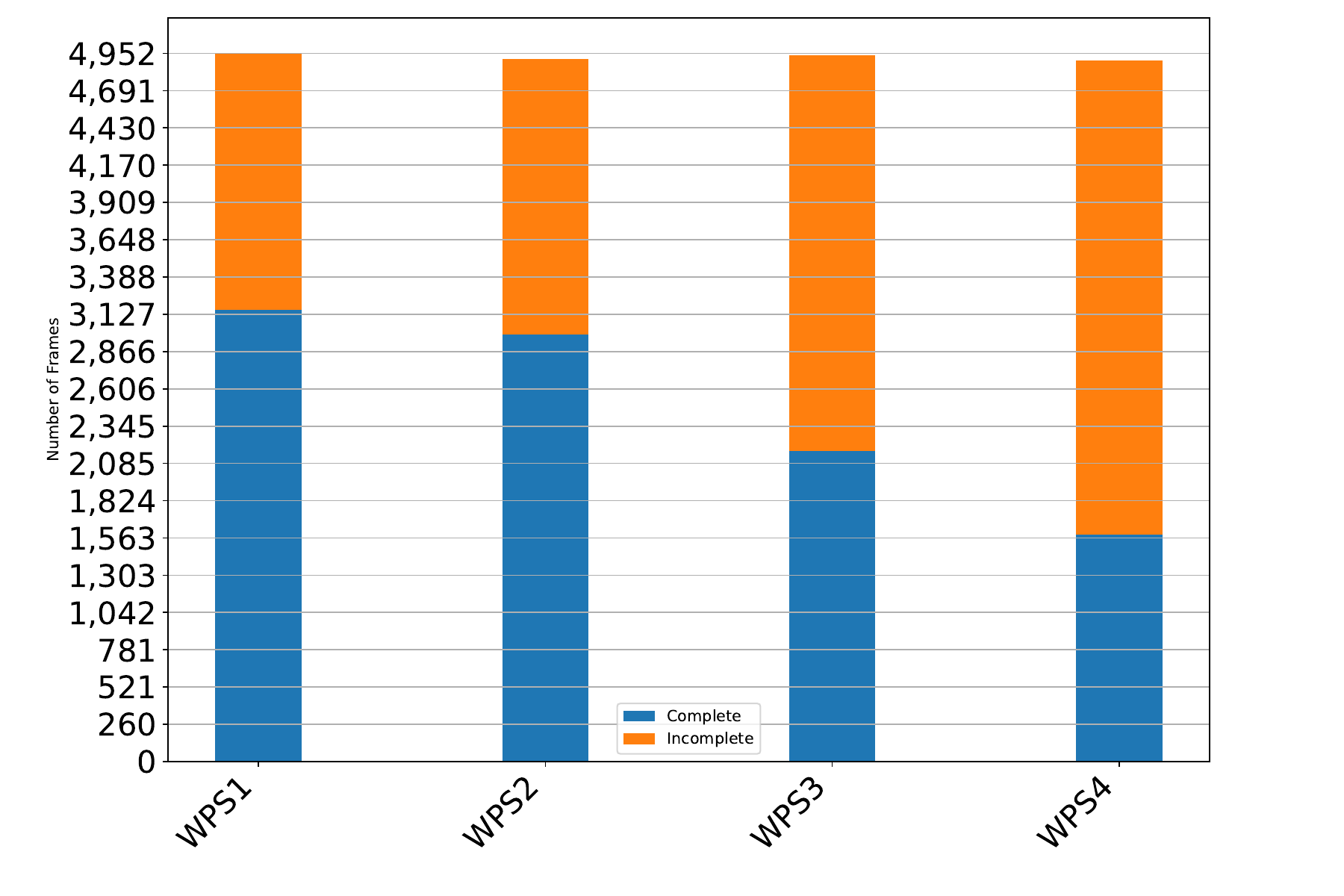}
    \caption{Task load variance}
    
    \label{fig:task_load_frame_Completion}
\end{subfigure}
\caption{Frame completion}
\end{figure}

The key metric of our experiments is whether or not algorithms are capable of processing frame pipelines from beginning to end successfully. This metric will be higher in experiments with lighter workloads. Under the highest load (weighted 4 workload) we can see in fig.  \ref{fig:preempt_vs_non_preempt_frame_Completion} that the scheduler outperforms the workstealers when using both preemption and non-preemption approaches. The scheduler is able to complete the most frames, completing 32.4\% of frames with preemption and 29.36\% without preemption. The preemption scheduler completes the highest amount of frames across scenarios, completing $5\%$ more than its non-preemption counterpart in the uniform workload ($50$ and $45$ percent). 
This is due to increased completion rate of \textit{high priority} tasks with preemption, as will be discussed later.
The decentralised workstealer has the worst performance with 8.96\% completed frames for the preemption mechanism and only 5.64\% frames completed in its non-preemption counterpart. 

The centralised workstealer slightly outperformed the preemption capable decentralised workstealer with 9.65\% completed frames for preemption and 9.23\% more for non-preemption. 
Overall while the decentralised workstealer performs slightly worse than its centralised counterpart, their performance remain comparable. The main reason for the decentralised workstealer's slight performance drop is the random access to resources, whenever the decentralised workstealer queries for a job it must query multiple devices in a random fashion until it finds a device with tasks available to process. This also results in more tasks processing without consideration of what set each low-priority request belongs to. In the non-preemption scenario this is not a critical issue, however, when preemption is introduced the preemption of low priority tasks with the random access nature of task allocation results in fewer tasks from the same set receiving placement and when they do receive placement processing to completion without experiencing preemption.

To understand how the schedulers' performance varies under different levels of task load we examine it under the weighted X configurations of increased network saturation. We can see in fig. \ref{fig:task_load_frame_Completion} under the variation of load that when the network is weighted to generate one or two low-priority tasks the number of frames completed remains comparable. This is to be expected as in most cases devices are able to process two low-priority tasks locally, with no need for offloading and preemption. However, as the workload becomes weighted toward generating three low priority tasks per set the number of completed frames drops by 16.71\% but when increasing the weighting of low priority tasks to four it only drops by 11.61\%. This reinforces that the critical point for performance drop is when capacity of local devices is exceeded forcing them to rely on other devices to process their excess workloads.

\subsection{Pipeline Stage Analysis}
\begin{figure}
    \centering
\begin{subfigure}{.49\textwidth}
    \includegraphics[width=\textwidth]{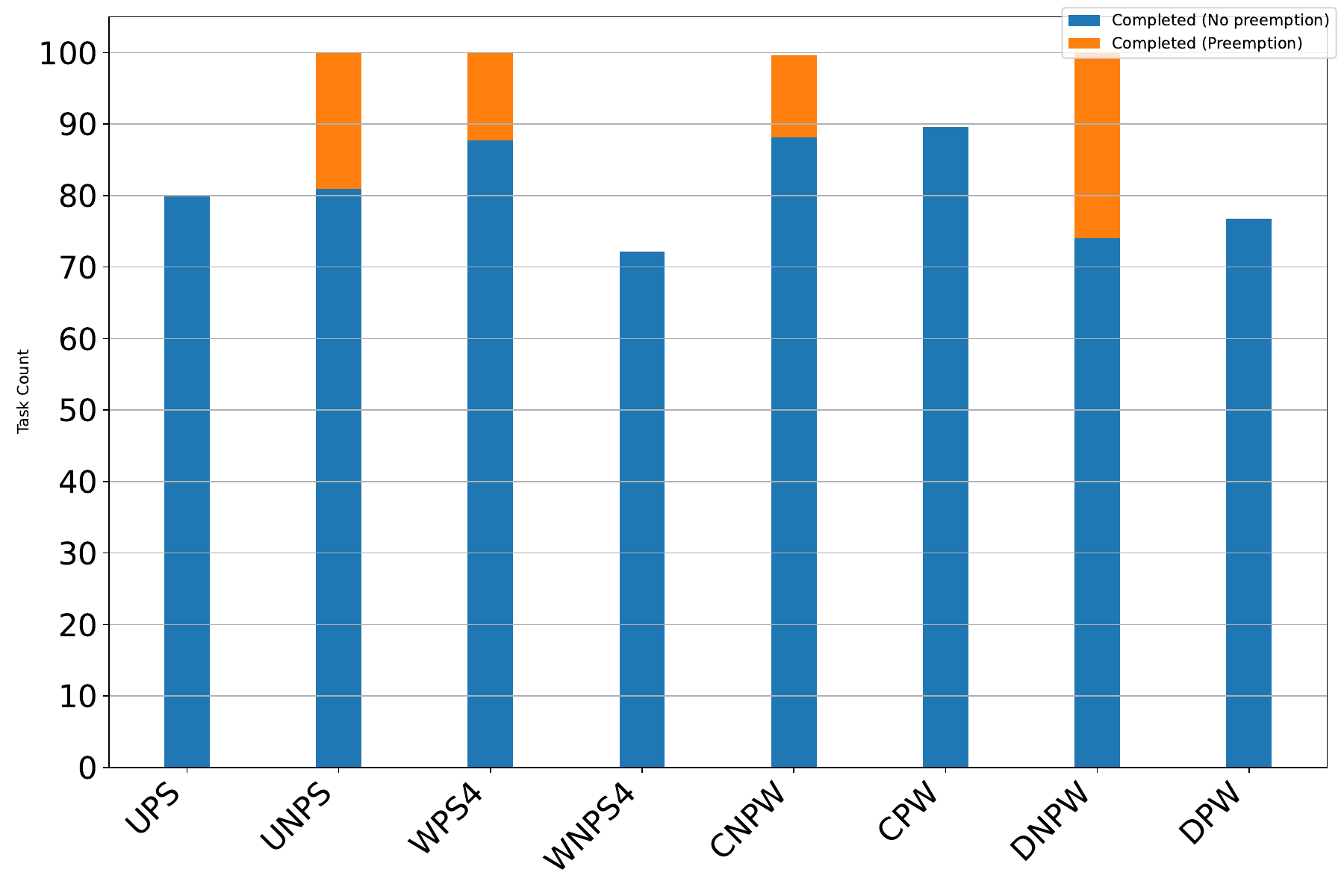}
    \caption{Preemption vs. non-preemption}
    \label{fig:preempt_vs_non_preempt_high_priority_low_comp_res}
\end{subfigure}
\begin{subfigure}{0.49\textwidth}
    \centering
    \includegraphics[width=\textwidth]{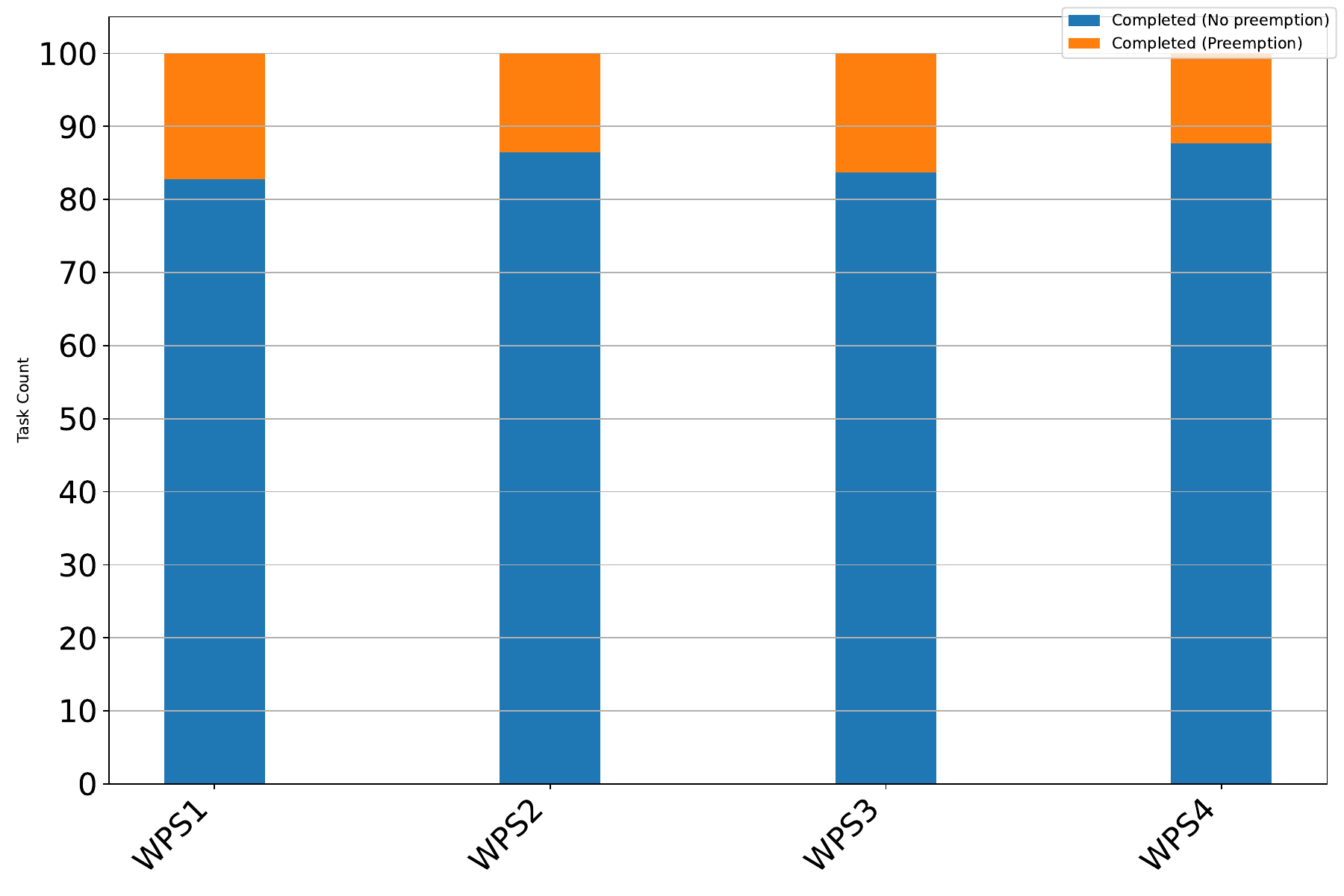}
    \caption{Task load variance}
    \label{fig:task_load_high_priority_low_comp_res}
\end{subfigure}
\caption{High-priority low-complexity completion}
\end{figure}

The direct impact of preemption can be seen in Fig. \ref{fig:preempt_vs_non_preempt_high_priority_low_comp_res} and \ref{fig:task_load_high_priority_low_comp_res}. The high priority completion rate shows us how many high priority tasks are completed as a percentage of the total amount of high priority tasks generated. This figure details the amount of high priority tasks that completed and those that had to preempt low priority tasks to receive network resources in scenarios with preemption mechanisms. 

Preemption awareness results in $99\%$ of high priority tasks receiving resources they need, the 1\% drop can be attributed to runtime performance deviations. Under the uniform scenario the non-preemption scheduler completes 80\% of all high priority tasks. In the weighted four scenario we can see the non-preemption scheduler complete 72.1\% of all high priority tasks. The workstealers outperform the scheduler in non-preemption scenarios with the centralised workstealer completing 89.56\% and the decentralised workstealer completing 76.75\% of all high priority tasks. This in part may be due to the fact that workstealers perform better when allocating tasks that have no dependencies to other tasks.

We observe that when the number of high priority tasks completed increase there is a corresponding increase in the total amount of frames completed. This can be seen in Fig. \ref{fig:preempt_vs_non_preempt_high_priority_low_comp_res} and \ref{fig:preempt_vs_non_preempt_frame_Completion} in the uniform scenario where the preemption capable scheduler completes $20\%$ more high priority tasks than the non-preemption version which corresponds to 224 more frames completed. This is because more frames are entering the system due to preemption, additionally, in many frames they only require the object detector and high priority stage without generating any low priority sets of their own. 
Both schedulers under the uniform scenario complete roughly $80\%$ without requiring preemption.

The preemption aware decentralised workstealer performs the worst in preemption scenarios with $74\%$ without invoking preemption, this in part is due to the uncoordinated nature of workstealers combined with the increased saturation experienced under preemption when high priority tasks are able to generate low priority workloads that would not previously enter the network.

\begin{figure}
    \centering
\begin{subfigure}{.49\textwidth}
    \includegraphics[width=\textwidth]{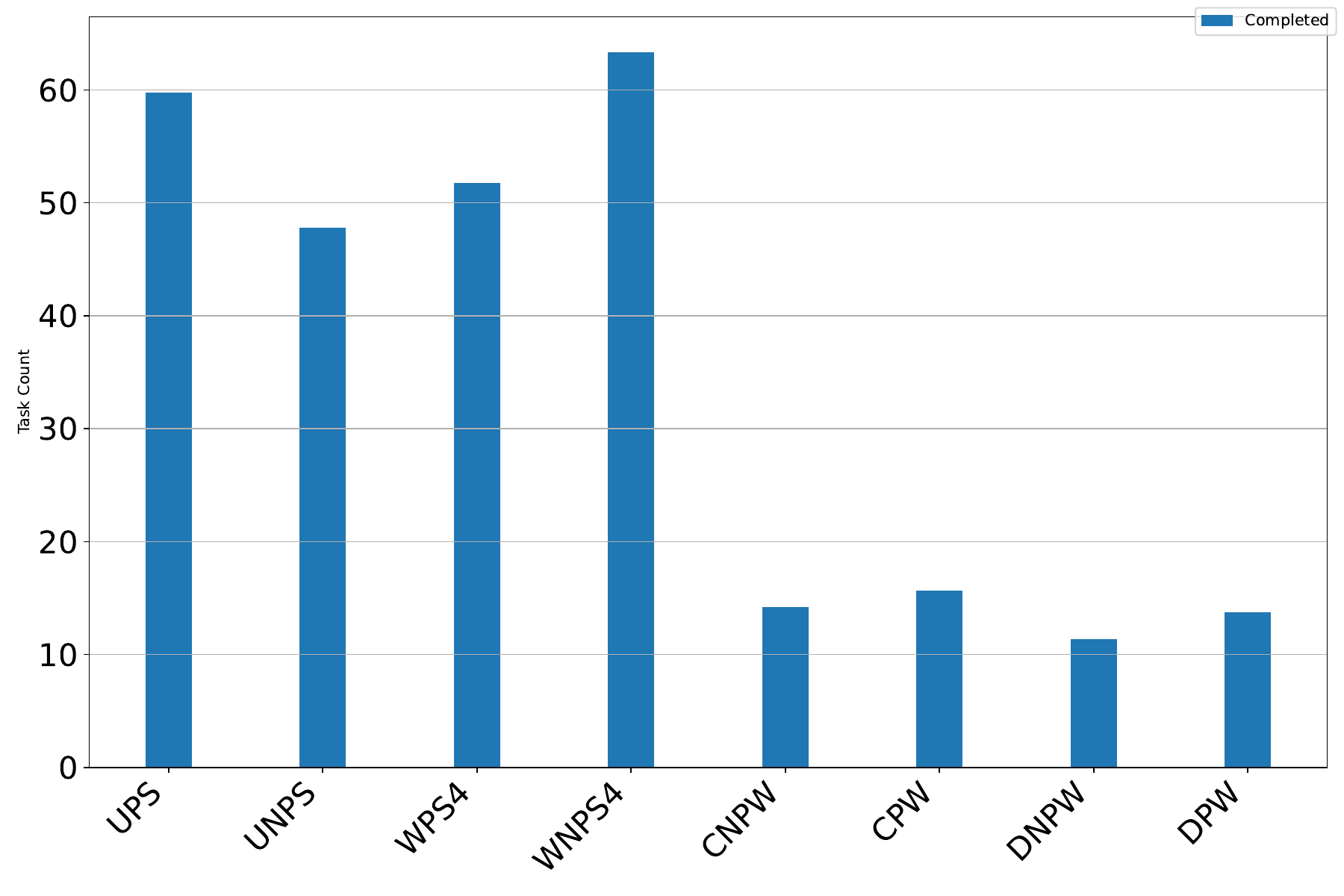}
    \caption{Preemption vs. non-preemption}
    \label{fig:preempt_vs_non_preempt_low_priority_high_comp}
\end{subfigure}
\begin{subfigure}{.49\textwidth}
    \centering
    \includegraphics[width=\textwidth]{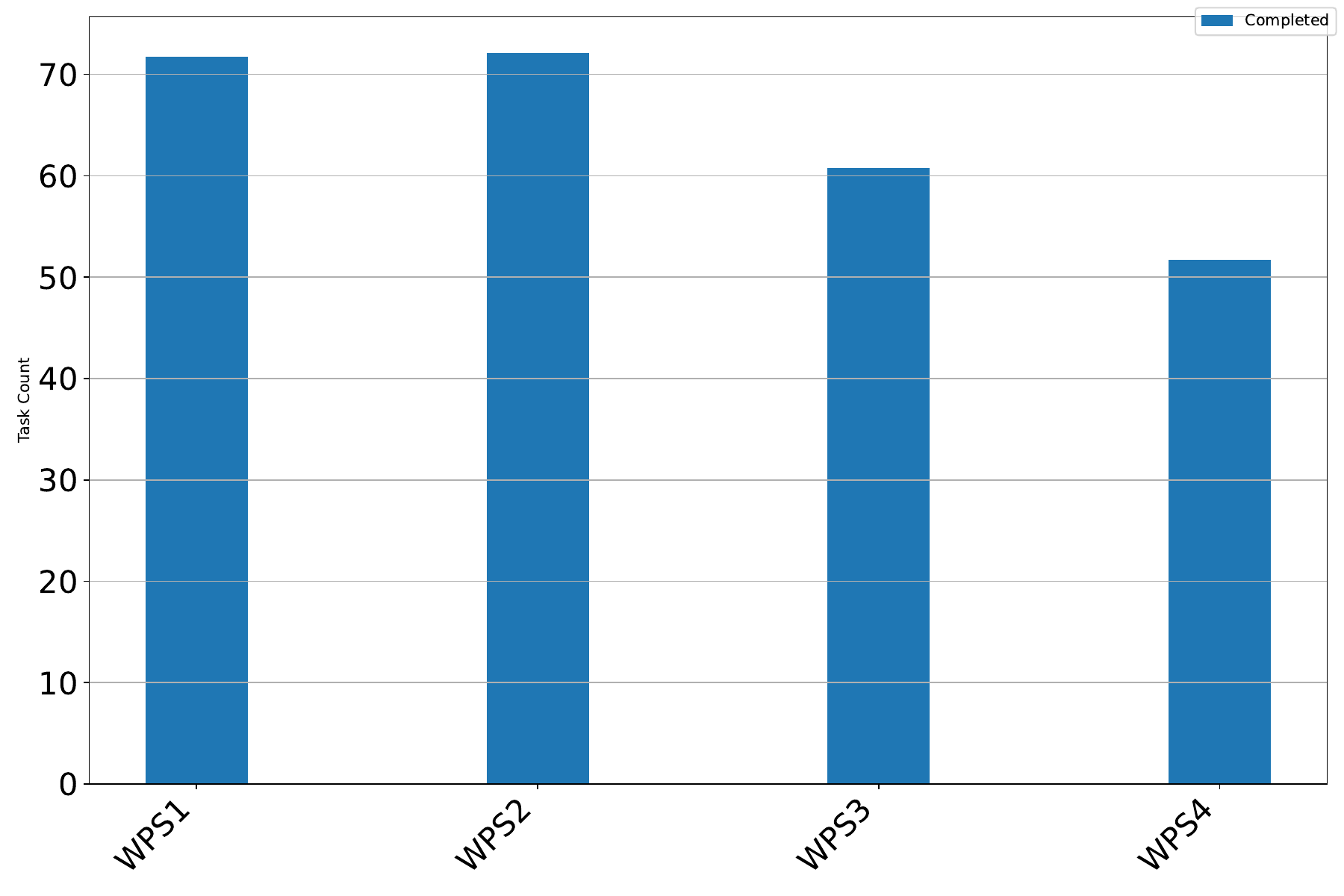}
    \caption{Task load variance}
    \label{fig:task_load_low_priority_high_comp}
\end{subfigure}
\caption{Low-priority completion by mechanism}
\end{figure}

Fig. \ref{fig:preempt_vs_non_preempt_low_priority_high_comp} and \ref{fig:task_load_low_priority_high_comp} show us the raw task completion value for low priority tasks by scenario and mechanism. 
We observe the negative impact of preemption on low-priority tasks in the uniform scenario, for individual low-priority task completion rates, the non-preemption scheduler completes 11.95\% more of its generated low-priority tasks compared its preemption counterpart.

In the weighted scenarios as the load increases the performance begins to drop, with the results for weighted one to four as follows: 71.71\%, 72.07\%, 60.78\% and 51.73\%. The main impact on performance is the jump from weighted two to three where offloading becomes more frequent as requests are more likely to contain more tasks than can be processed locally. %
Comparing the schedulers and the workstealers in the weighted four scenario we see that the schedulers complete more of their generated low priority tasks with 51.73\% and 63.31\% for preemption and non-preemption compared to the centralised preemption 15.65\% and non-preemption 13.76\% and finally the decentralised 14.20\% and 11.36\%. Overall, the non-preemption schedulers complete a higher percentage of low priority tasks generated than their preemption counterpart, driven by high-priority tasks preempting to receive resources. However, table  \ref{tab:low_priority_gen} shows that preemption scenarios generate far more low priority tasks than their non-preemption equivalent, as a result non-preemption completes a higher percentage of generated low-priority tasks whereas preemption complete a higher volume.

\begin{table}[]
\begin{tabular}{l|l}
\textbf{Scenario}                                   & \textbf{Low Priority Tasks Generated} \\ \hline
Uniform Scheduler (Preemption)             & 8640                                  \\ \hline
Uniform Scheduler (Non-Preemption)         & 6961                                  \\ \hline
Weighted 1 Scheduler (Preemption)         & 9296                                  \\ \hline
Weighted 2 Scheduler (Preemption)          & 10372                                 \\ \hline
Weighted 3 Scheduler (Preemption)          & 12973                                 \\ \hline
Weighted 4 Scheduler (Preemption)          & 13941                                 \\ \hline
Weighted 4 Scheduler (Non-Preemption)      & 9966                                  \\ \hline
Centralised Workstealer (Preemption)       & 13800                                 \\ \hline
Centralised Workstealer (Non-Preemption)   & 12414                                 \\ \hline
Decentralised Workstealer (Preemption)     & 13935                                 \\ \hline
Decentralised Workstealer (Non-Preemption) & 10671                                
\end{tabular}%
\caption{Total low priority tasks generated in each scenario.}
\label{tab:low_priority_gen}
\end{table}

\begin{figure}
    \centering
\begin{subfigure}{.49\textwidth}
    \includegraphics[width=\textwidth]{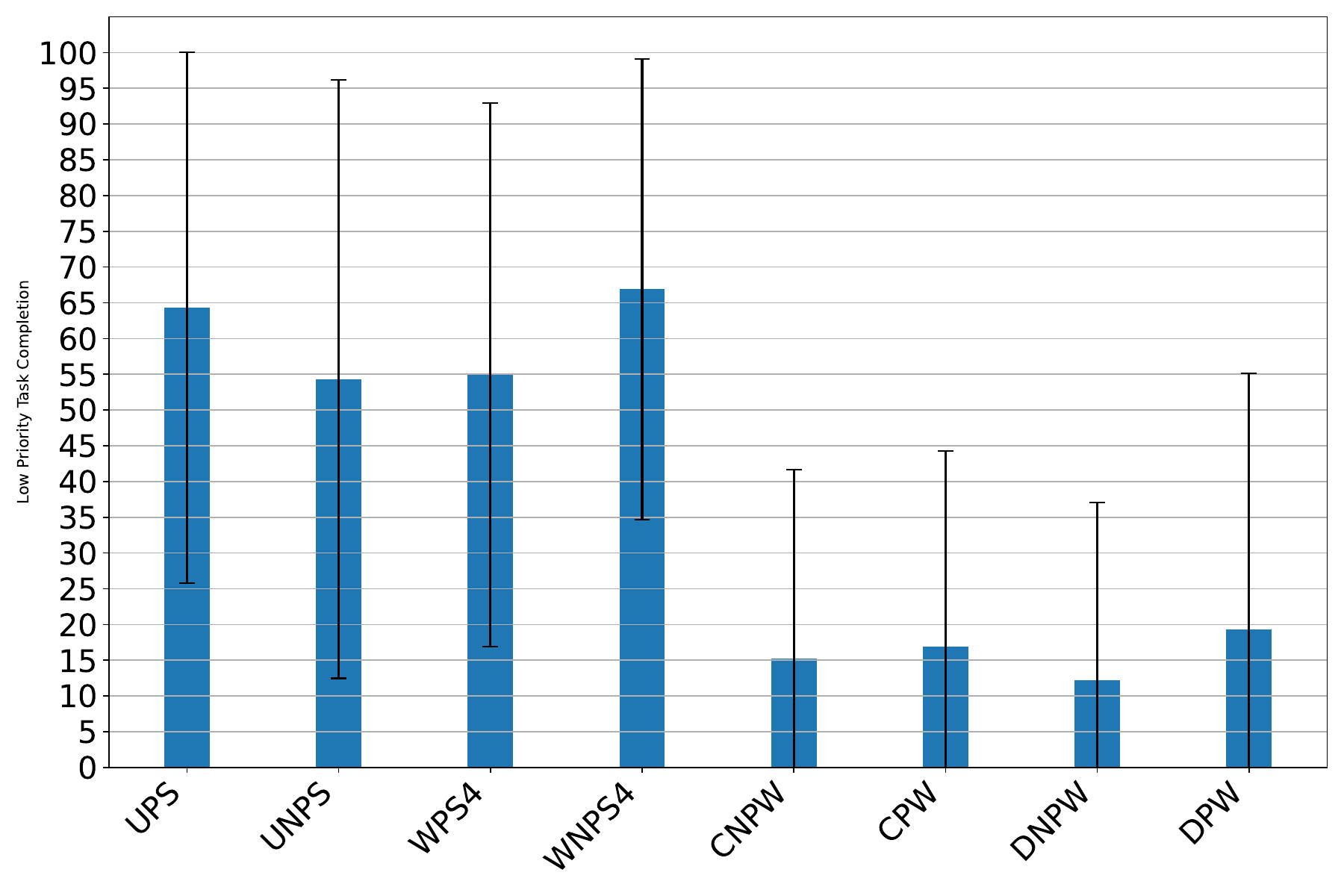}
    \caption{Preemption vs. non-preemption}
    \label{fig:preempt_vs_non_preempt_avg_low_priority_per_request}
\end{subfigure}
\begin{subfigure}{.49\textwidth}
    \includegraphics[width=\textwidth]{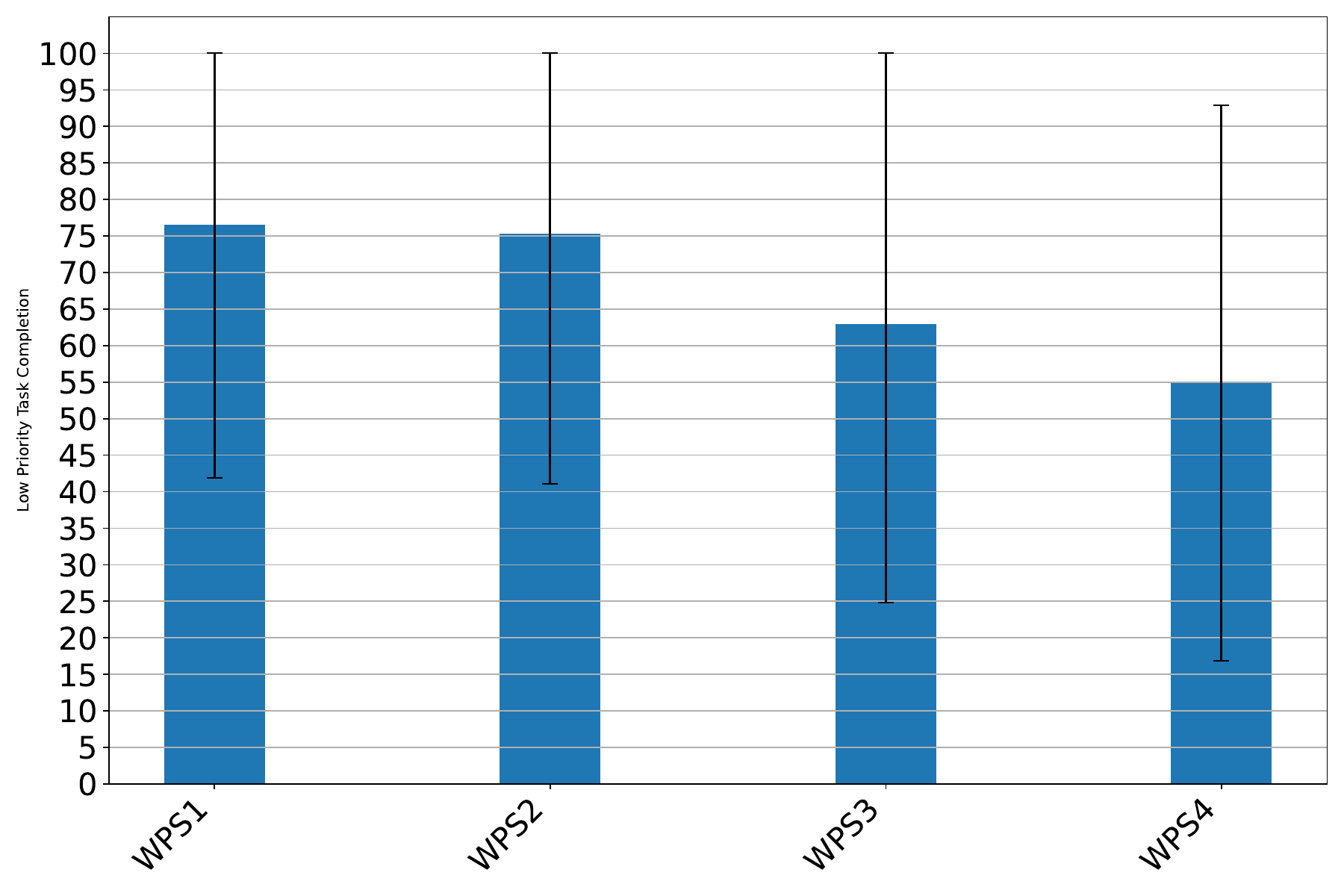}
    \caption{Task load variance}
    \label{fig:task_load_avg_low_priority_per_request}
\end{subfigure}
\caption{Low-priority task completion per request.}
\end{figure}

A low priority request isn't considered complete unless the full task set has been allocated and successfully executed. While some approaches such as the scheduler allocate low-priority tasks in sets, none of the solutions presented prioritise the completion of task sets over individual low-priority tasks.  The percentage of low priority tasks completed per request seen in fig. \ref{fig:preempt_vs_non_preempt_avg_low_priority_per_request} and \ref{fig:task_load_avg_low_priority_per_request} allow us to see how effective an algorithm is at actually satisfying the goal of completing frames versus simply completing tasks. As can been seen in the figure, algorithms that implement a preemption mechanism have a lower result on average, this a result of offloaded tasks occupying resources resulting in preemption with failed reallocations. %
We can see that when both are given a uniform workload the no-preemption scheduler completes $10\%$ more tasks per low priority request than their preemption counterpart. Due to the myopic nature of their approaches, the workstealers perform much worse with $23\%$ at their highest (decentralised with no preemption) and $15\%$ at their lowest (centralised with preemption). Finally, we observe in fig. \ref{fig:task_load_avg_low_priority_per_request} when given a weighted workload the preemption scheduler completes roughly $\sim75\%$ in weighted 1 and 2. However, once offloading becomes more frequent the performance drops by  $\sim$10\% per load increase. 

\begin{figure}
    \centering
\begin{subfigure}{.49\textwidth}
    \includegraphics[width=\textwidth]{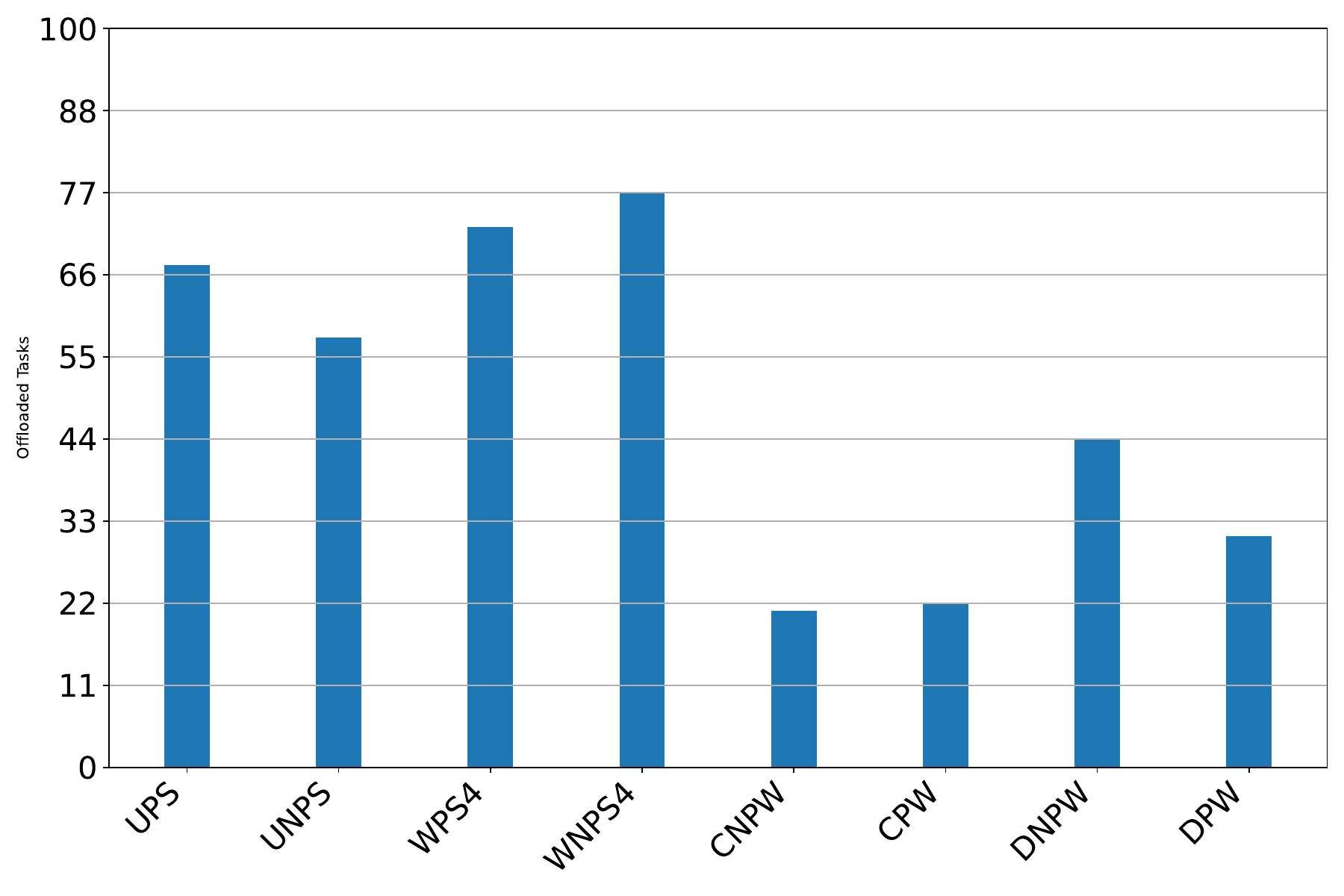}
    \caption{Preemption vs. non-preemption}
    \label{fig:preempt_vs_non_preempt_offload_completion}
\end{subfigure}
\begin{subfigure}{.49\textwidth}
    \includegraphics[width=\textwidth]{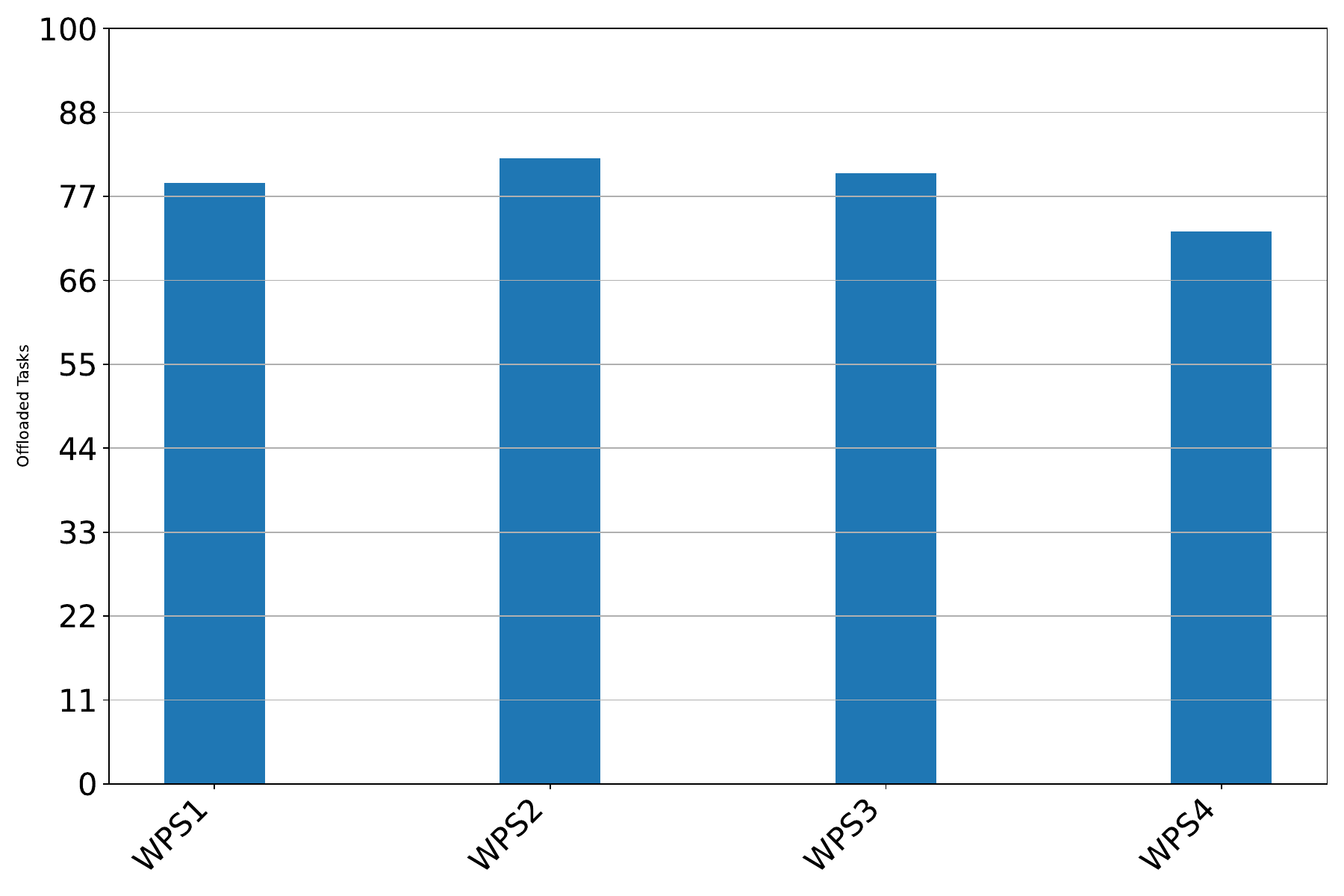}
    \caption{Task load variance}
    \label{fig:task_load_offload_completion}
\end{subfigure}
\caption{Offloaded high-complexity low-priority task completion rate by mechanism}
\end{figure}

Overall we observe from fig. 
\ref{fig:preempt_vs_non_preempt_low_priority_high_comp}, 
\ref{fig:task_load_low_priority_high_comp},
\ref{fig:preempt_vs_non_preempt_avg_low_priority_per_request}, \ref{fig:task_load_avg_low_priority_per_request}, \ref{fig:preempt_vs_non_preempt_offload_completion} and \ref{fig:task_load_offload_completion} that preemption allows more high priority tasks to enter the network which generate low priority tasks of their own; the increased network saturation means that many of these additional tasks both local and offloaded do not get to complete processing; resulting in an offloaded completion count comparable to their non-preemption counterparts even if the total percentage of offloaded tasks completed is lower.

\begin{figure}
    \centering
\begin{subfigure}{.49\textwidth}
    \includegraphics[width=\textwidth]{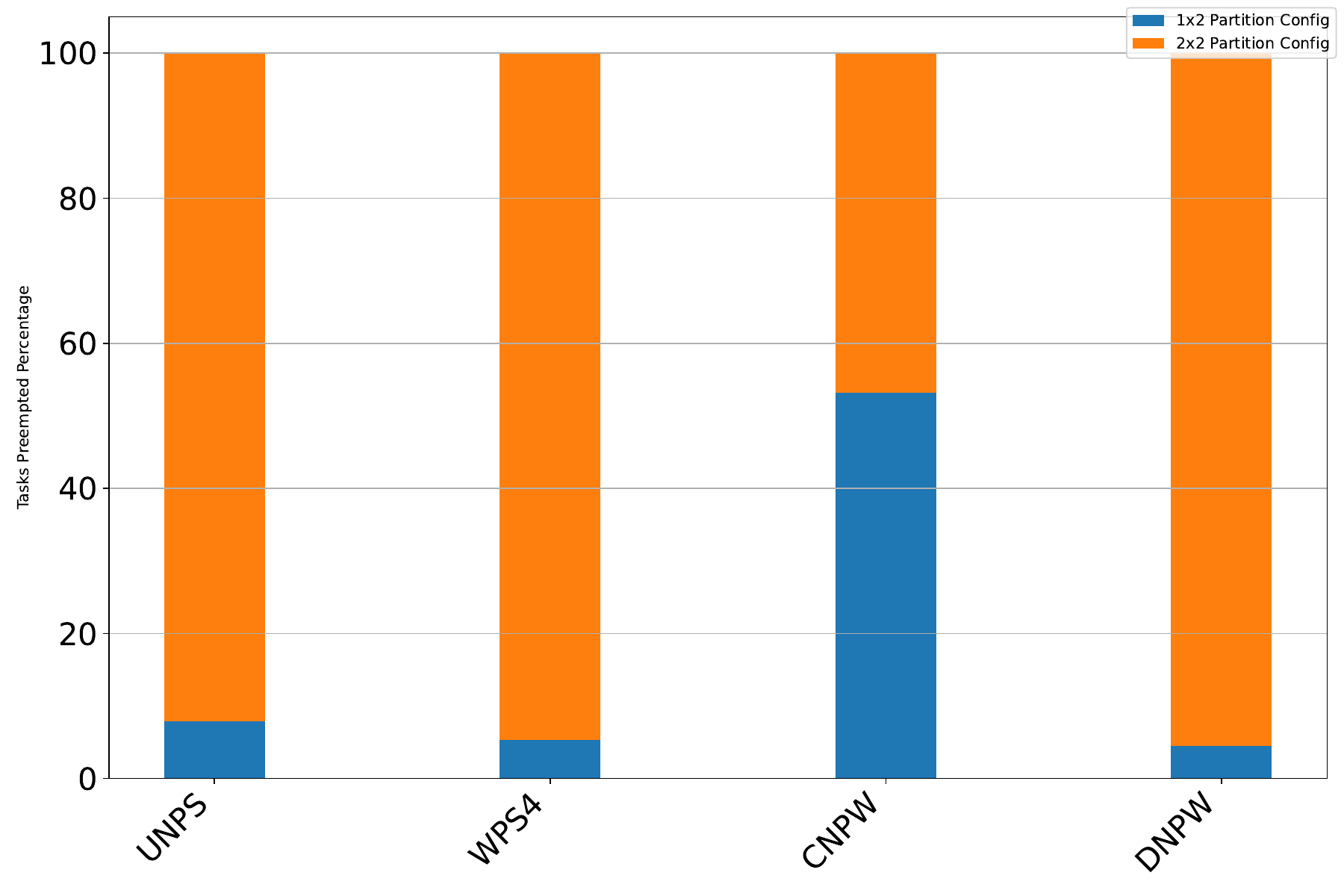}
    \caption{Preemption vs. non-preemption}
    \label{fig:preempt_vs_non_preempt_partition_config_preemption}
\end{subfigure}
\begin{subfigure}{.49\textwidth}
    \centering
    \includegraphics[width=\textwidth]{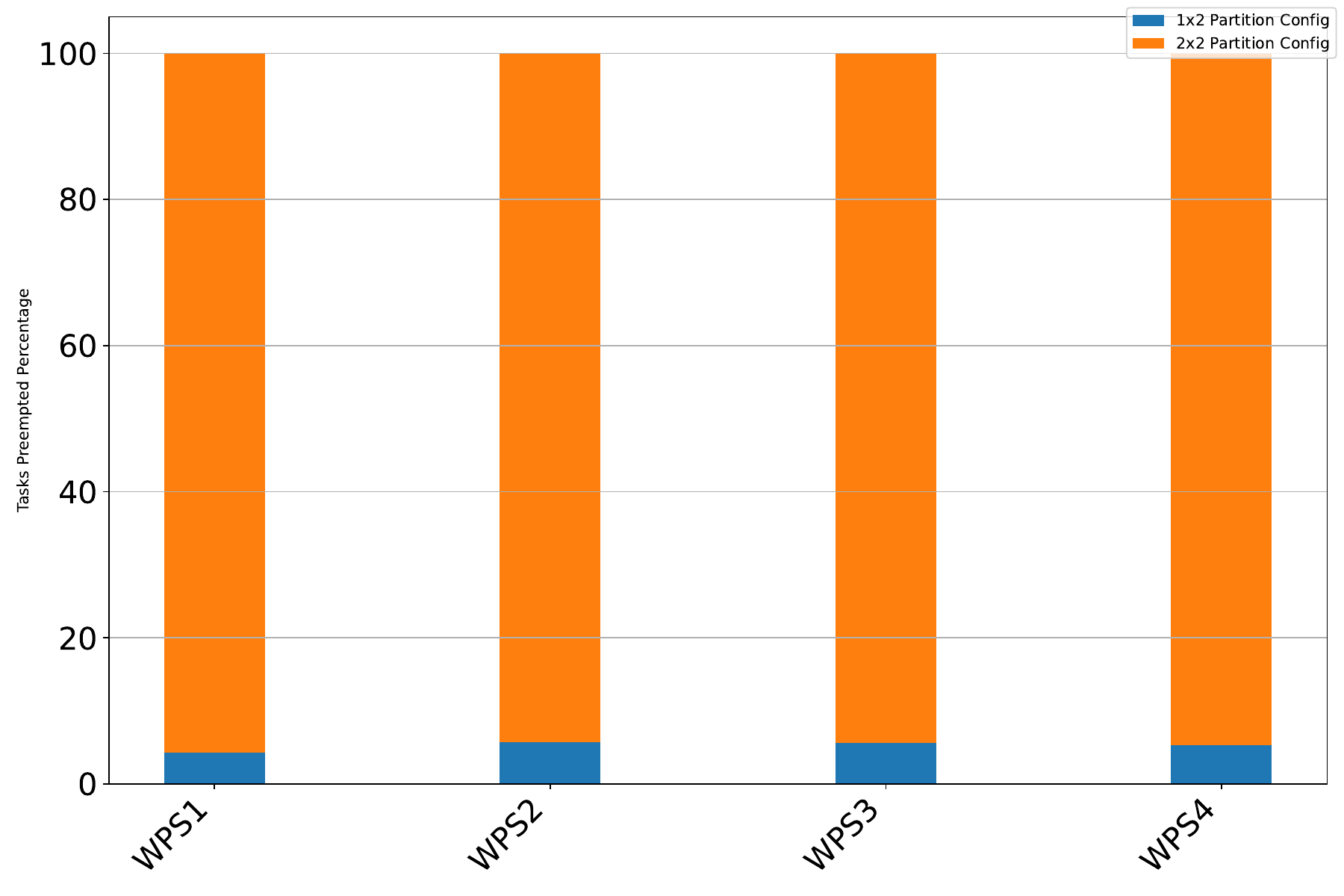}
    \caption{Task load variance}
    \label{fig:task_load_partition_config_preemption}
\end{subfigure}
\caption{Percentage of preempted tasks by partition configuration for the preemption mechanism}
\end{figure}

\begin{figure}
    \centering
    \includegraphics[width=.5\textwidth]{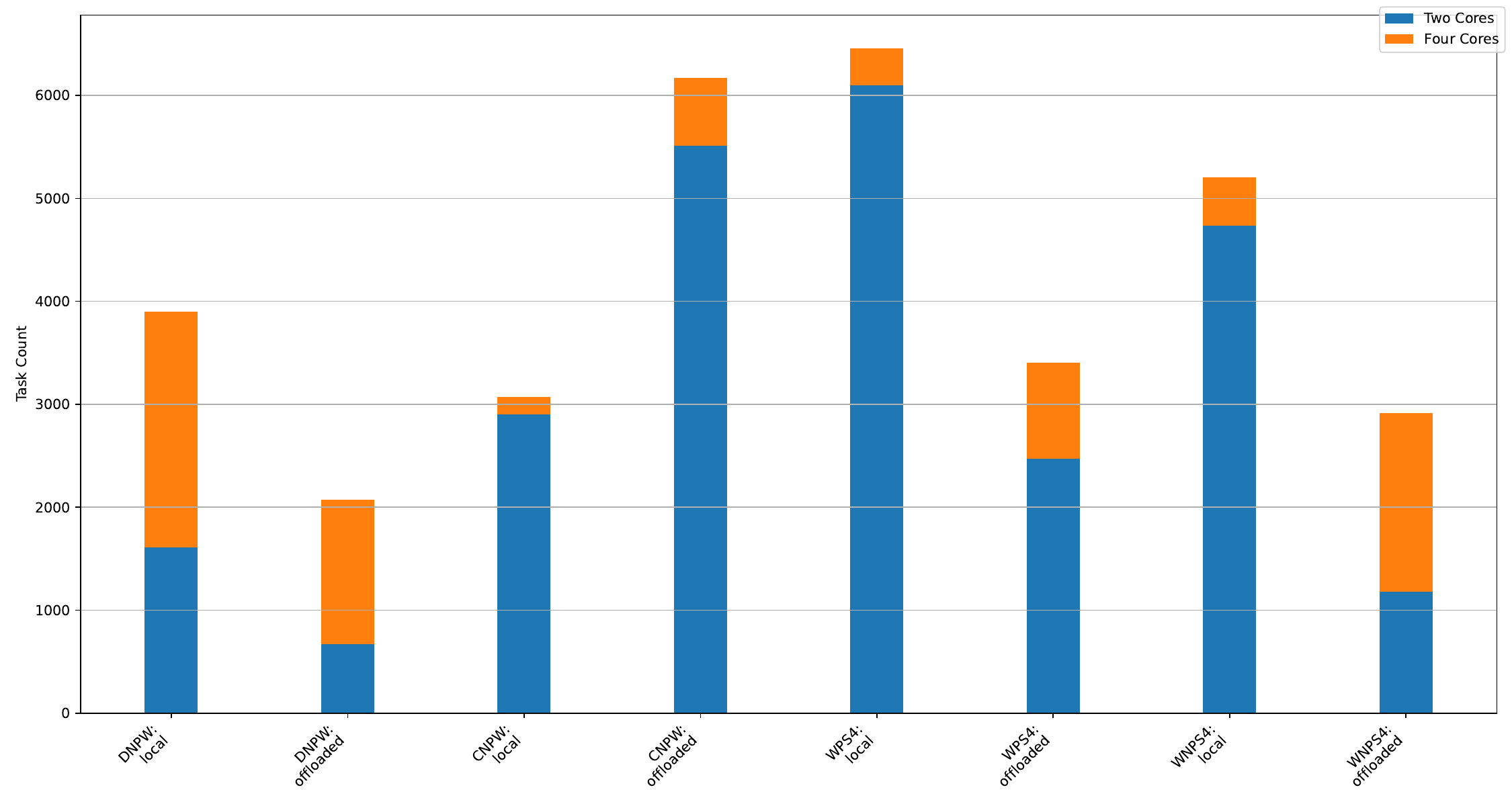}
    \caption{Core Allocation of local and offloaded tasks across four experiments.}
    \label{fig:core_allocation}
\end{figure}

In Fig. \ref{fig:preempt_vs_non_preempt_partition_config_preemption} and \ref{fig:task_load_partition_config_preemption} we observe the configuration of preempted low-priority tasks. 
We see in most scenarios that on average a task is more likely to experience preemption when it fully occupies the resources of a device.

Fig. \ref{fig:core_allocation} shows the core distribution of allocated low priority tasks both local and offloaded across all solutions under the weighted 4 scenario. In most scenarios, tasks are allocated two-cores. As mentioned previously the scheduler attempts to pack in tasks locally which is why the scheduler's local tasks skew towards two-core allocations in every scenario. However, when performing offloading the scheduler always tries to increase the resources allocated to low-priority tasks. Fig. \ref{fig:preempt_vs_non_preempt_partition_config_preemption} shows the high preemption rate for four-core allocations indicates that a more conservative approach to resource partitioning could be beneficial. By allocating additional resources only when there is a risk of missing deadlines, the scheduler could reduce the need for preemption while still meeting all deadlines.

We observe in table \ref{tab:preemption_statistics} the success rate of preempted tasks that are scheduled for reallocation across each preemption scenario (with the exception of the centralised workstealer as we were unable to collect reallocation statistics for it). The weighted 2 scenario contains the best performance where only 2 tasks successfully receive reallocation. Across all the listed scenarios, when preemption occurs, it is extremely unlikely that the task will receive reallocation successfully. Due to the experimental setup, our devices process in staggered pairs. Preemption typically occurs after a significant amount of time has been spent on processing the preempted task. Combined with high reallocation times as will be discussed in \ref{sec:processing_times}, there is little time left to attempt reallocation before the task deadline.

\begin{table}[]
\begin{tabular}{l|ll}
\textbf{Scenario}                               & \textbf{Preemption Failure} & \textbf{Preemption Success} \\ \hline
\textbf{Uniform Scheduler (Preemption)}         & 822                         & 1                           \\ \hline
\textbf{Weighted 1 Scheduler (Preemption)}      & 855                         & 0                           \\ \hline
\textbf{Weighted 2 Scheduler (Preemption)}      & 664                         & 2                           \\ \hline
\textbf{Weighted 3 Scheduler (Preemption)}      & 807                        & 0                           \\ \hline
\textbf{Weighted 4 Scheduler (Preemption)}      & 601                         & 1                           \\ \hline
\textbf{Decentralised Workstealer (Preemption)} & 1256                        & 1                          
\end{tabular}%
\caption{Post-preemption reallocation for low-priority tasks.}
\label{tab:preemption_statistics}
\end{table}

\subsection{Processing Times}
\label{sec:processing_times}
The processing time of the high-priority scheduling algorithm is proportional to the number of tasks that have been allocated to the source device. In the scenario where there are not enough resources locally to execute the high-priority task, the act of selecting a low-priority task to preempt is also proportional to the number of tasks that have been allocated to the source device. Therefore, the complexity of a preemption scenario is  $O(3 * number\_of\_local\_tasks)$.

Low-priority tasks are not restricted to a single device. If no resources are available during the current timeslot, the scheduler will iterate through the completion times of tasks across the network as it searches again. Therefore, the processing time to allocate low-priority tasks is proportional to the $O(number\_of\_tasks^{2})$ where number of tasks is the total number of tasks allocated in the network. When a low-priority task is preempted we consider it's reallocation time to be the time from which it is preempted till it has received reallocation, it is proportional to the $O(number\_of\_local\_tasks + number\_of\_tasks^{2})$
The overall deadline of the high-priority stage is quite low ($\sim$1 second) therefore the time it takes to allocate a high priority task can have a large impact on the performance of an approach. 

\begin{figure}
    \centering
\begin{subfigure}{.49\textwidth}
    \includegraphics[width=\textwidth]{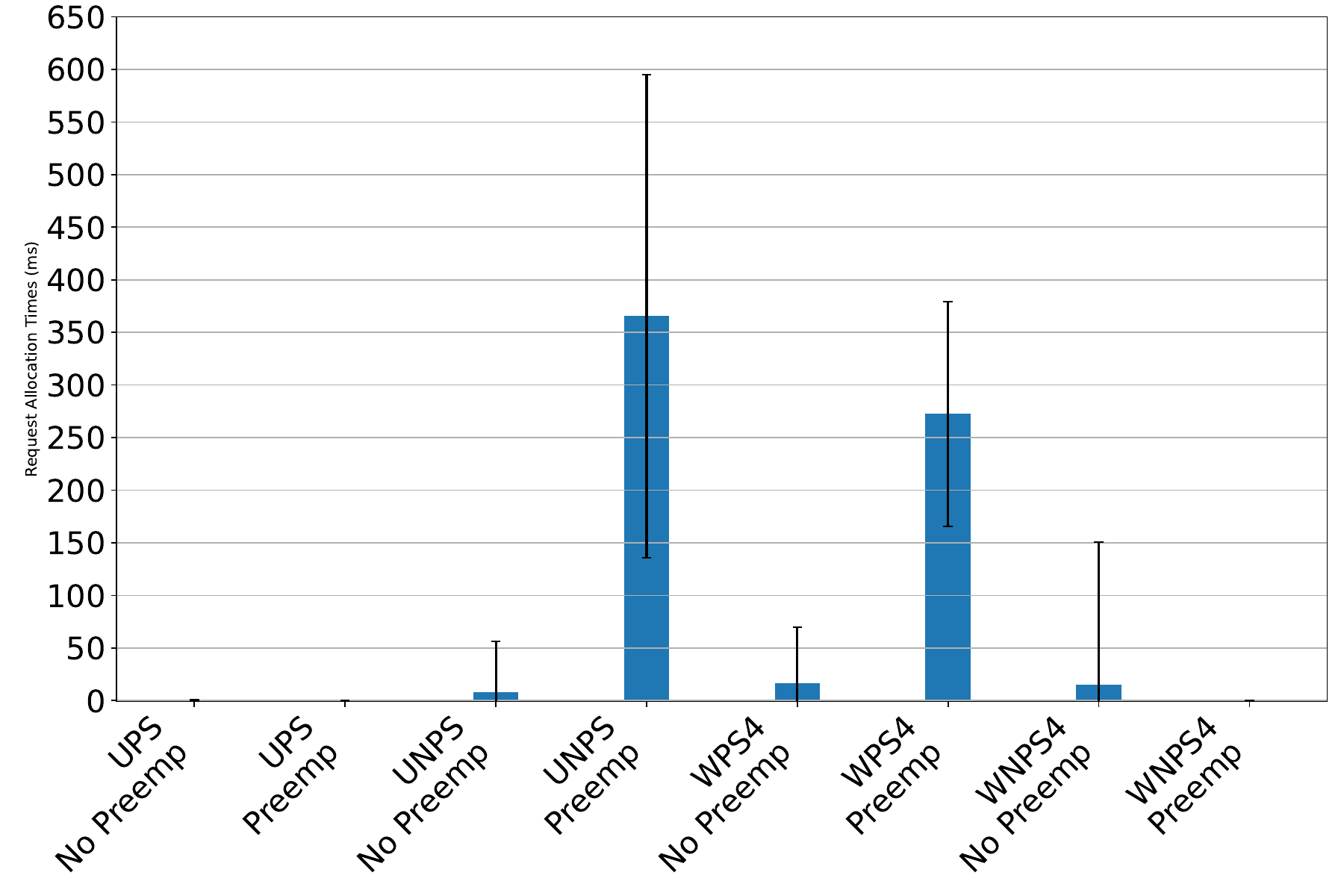}
    \caption{Preemption vs. non-preemption}
    \label{fig:preempt_vs_non_preempt_low_comp_system_time}
\end{subfigure}
\begin{subfigure}{.49\textwidth}
    \centering
    \includegraphics[width=\textwidth]{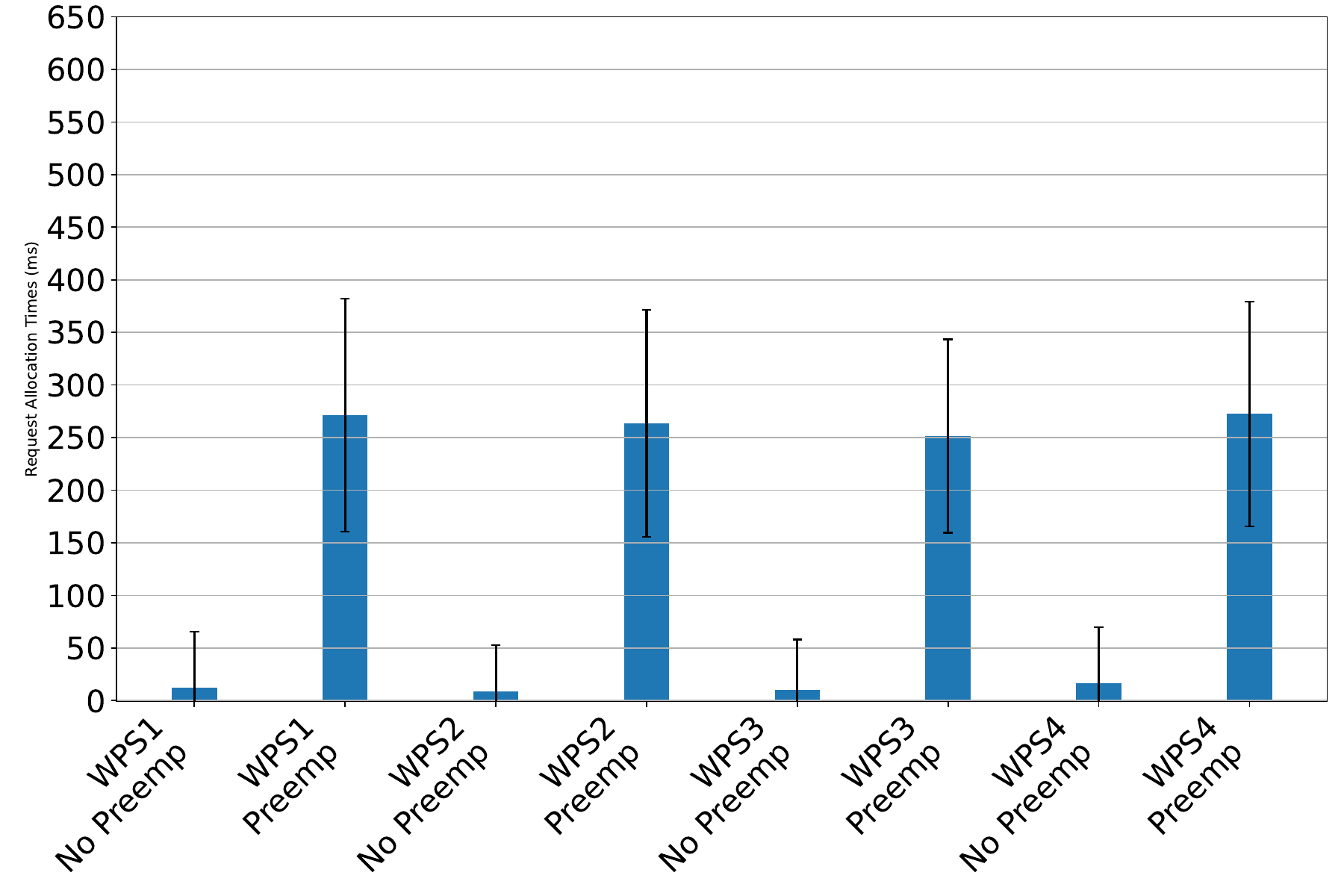}
    \caption{Task load variance}
    \label{fig:task_load_low_comp_system_time}
\end{subfigure}
\caption{Average low-complexity high-priority allocation time by mechanism}
\end{figure}

We can see both the time it takes for an initial allocation of a high priority task and a high priority task that has invoked preemption in Fig. \ref{fig:preempt_vs_non_preempt_low_comp_system_time} and \ref{fig:task_load_low_comp_system_time}. Higher values are representative of a more heavily loaded network (either due to a higher workload or because the approach is able to allocate more tasks into the network). As our devices cannot generate high and low priority tasks simultaneously and high priority tasks can only allocate to their source device, a larger value also indicates that the scheduler is performing offloading more frequently. This increases high-priority search time and shows that a particular approach is distributing its workloads across the network. The non-preemption aware scheduler allocates in under 1ms in a uniform scenario with the preemption aware approach achieving 8ms for an initial allocation and 365ms for reallocation. The weighted scenarios are as follows: weighted one initial 12.29ms and reallocation 271.52ms, weighted two initial 8.50ms and reallocation 263.42ms, weighted three initial 10.36ms and 251.43ms for reallocation. Even without invoking preemption, preemption aware approaches still suffer larger search times for allocating high-priority tasks. This stems from the increased amount of low-priority tasks in the network generated from other high-priority tasks on other devices that invoke preemption to process in scenarios where they previously would not receive processing.

\begin{figure}
    \centering
\begin{subfigure}{.49\textwidth}
    \includegraphics[width=\textwidth]{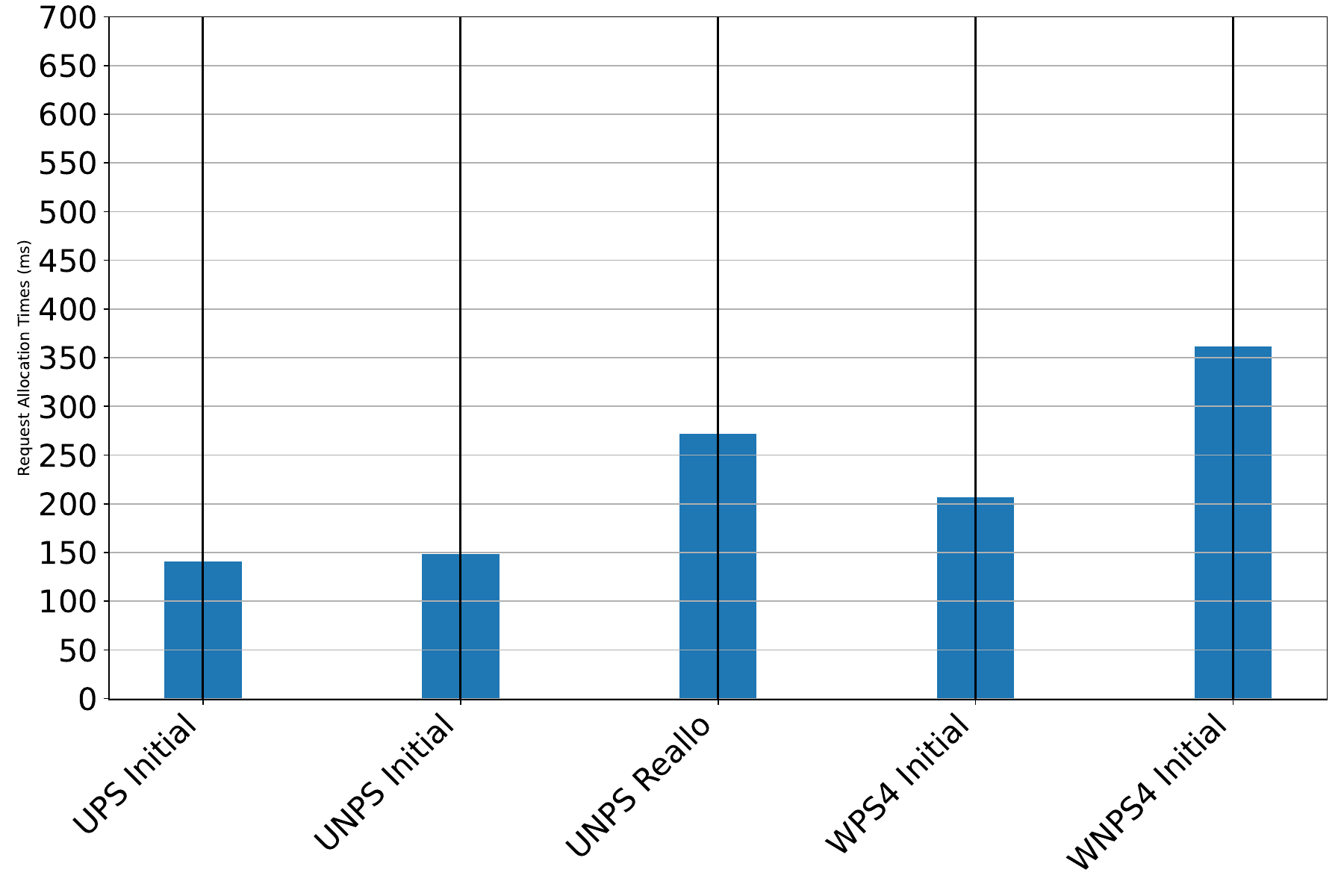}
    \caption{Preemption vs. non-preemption}
    \label{fig:preempt_vs_non_preempt_avg_high_comp_system_time}
\end{subfigure}
\begin{subfigure}{.49\textwidth}
    \includegraphics[width=\textwidth]{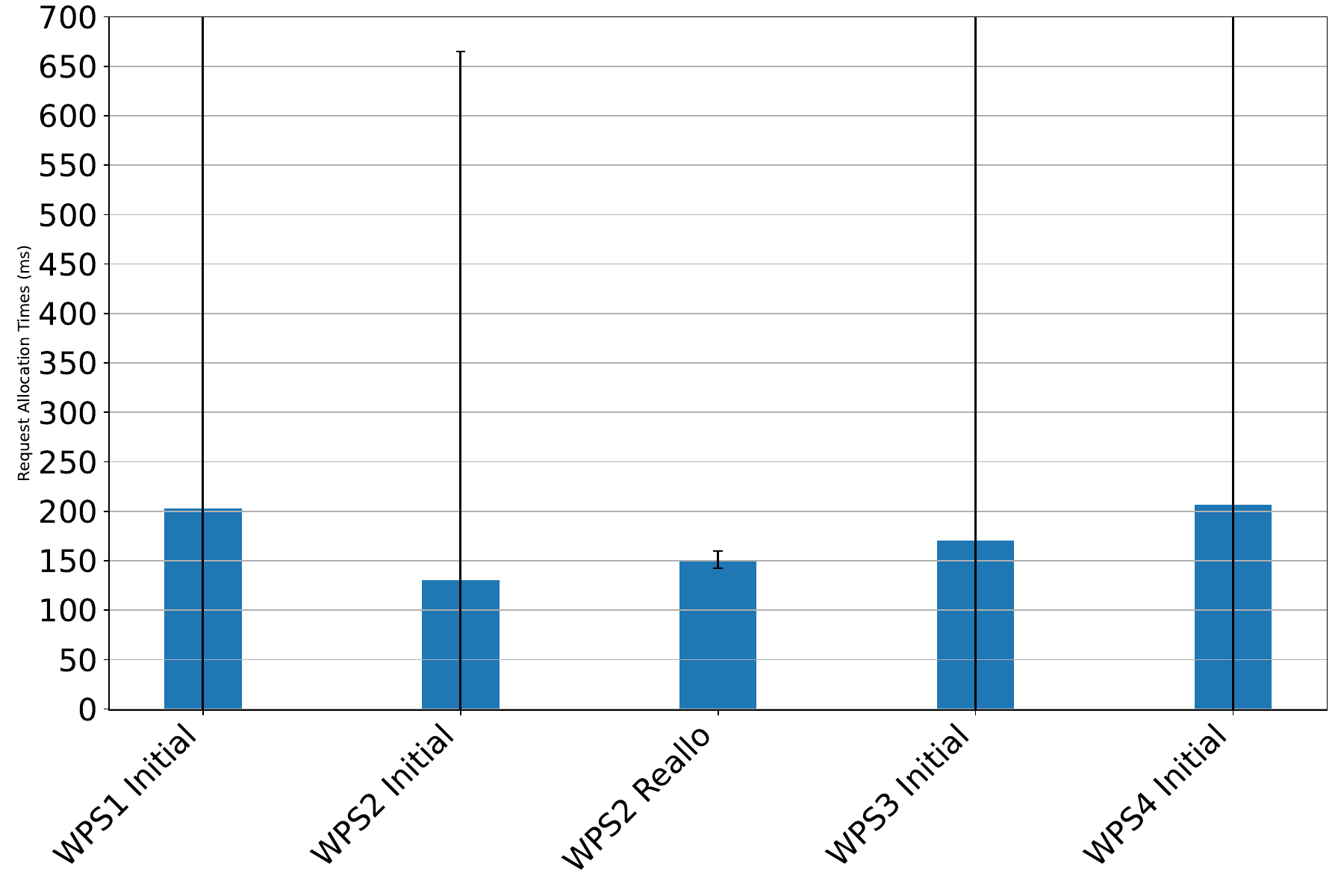}
    \caption{Task load variance}
    \label{fig:task_load_avg_high_comp_system_time}
\end{subfigure}
\caption{Average time to allocate high-complexity low-priority tasks}
\end{figure}

We observe across fig. \ref{fig:preempt_vs_non_preempt_low_comp_system_time}, \ref{fig:task_load_low_comp_system_time}, \ref{fig:preempt_vs_non_preempt_avg_high_comp_system_time} and \ref{fig:task_load_avg_high_comp_system_time} that the most significant impact network saturation has on system latency is the preemption low-priority reallocation scenario.
The results show that in a uniformly distributed network that the non-preemption mechanism incurs $150ms$ of scheduling latency with the preemption scheduler incurring a slightly lower allocation time of $148ms$ with no successful reallocations. The lack of successful reallocations is expected, one of the main bottlenecks in the low priority allocation algorithm is the search through task finish time-points for a future allocation. We consider reallocation time as the the time window between the point of task preemption and the final placement decision (reallocation or task cancellation) as this is the delay a task experiences following preemption. When preemption occurs, it is likely that the low priority preempted task was already allocated one of the few viable resources it could process in, this may force it to iterate through the finish times of tasks in the network (including the task that just preempted it) in order to find a viable mapping. We can see from 
 \ref{tab:preemption_statistics} that the limited availability of viable slots to process in is further reduced by the high latency incurred, making reallocation unlikely. Therefore, if preemption is to be utilised while still allowing for the possibility of reallocation then a light weight network model is required to reduce the impact a heavily loaded network has on search times.

\begin{table}[]
\resizebox{\columnwidth}{!}{%
\begin{tabular}{l|lll}
\textbf{Trace File} & \textbf{Potential Low Priority Task Count} & \textbf{Potential High Priority Task Count} & \textbf{Frames} \\ \hline
\textbf{Uniform}       & 8640  & 4320 & 1296 \\ \hline
\textbf{Weighted 1}    & 9296  & 4952 & 1296 \\ \hline
\textbf{Weighted 2}    & 10372 & 4915 & 1296 \\ \hline
\textbf{Weighted 3}    & 12973 & 4939 & 1296 \\ \hline
\textbf{Weighted 4}    & 13941 & 4901 & 1296 \\ \hline
\textbf{Network Slice} & 1018  & 362  & 96  
\end{tabular}%
}
\caption{Potential Task Count by Experiment Input Scenario}
\label{tab:trace_file_stats}
\end{table}

\section{Discussion}
\label{sec:discussion}
The preemption mechanism allows 99\% of high-priority frames to receive the resources they need and complete. As we consider a pipeline task scenario, by completing more high priority tasks the system is able to generate more low-priority requests and allocate their DNN tasks into the network which in turn has the downstream impact completing more frames overall than a system without a preemption mechanism as can be seen across each experiment in Fig. \ref{fig:preempt_vs_non_preempt_frame_Completion}, \ref{fig:task_load_frame_Completion}, \ref{fig:preempt_vs_non_preempt_high_priority_low_comp_res} and \ref{fig:task_load_high_priority_low_comp_res}. 

While the preemption mechanism does impact the completion of offloaded low priority tasks, even in the worst case scenario (decentralised workstealer), this only results in a $\sim$16\% difference when compared to the non-preemption capable counterpart as seen in Fig. \ref{fig:preempt_vs_non_preempt_offload_completion} with the overall frame completion remaining 7\% higher in the preemption capable decentralised workstealer. 
This demonstrates that while there is a cost to preemption, it is not significant and is justified by the increased gains in frame completion even in the least favourable conditions.

Allowing more high-priority tasks to enter the network results in more low-priority tasks generated and processed which reduces network idling time. Reduced network idling time means that the probability of successful allocation decreases. The increased number of tasks leads to higher algorithm search times for the scheduler to find placement of both high and low priority tasks as can be seen in Fig. \ref{fig:preempt_vs_non_preempt_low_comp_system_time}, \ref{fig:task_load_low_comp_system_time},\ref{fig:preempt_vs_non_preempt_avg_high_comp_system_time}, and \ref{fig:task_load_avg_high_comp_system_time} which can affect the probability of successful allocation. Finally, a higher number of tasks will experience a higher number of preempted tasks. This can be seen under the uniform scenario no-preemption scheduler and preemption scheduler complete roughly the same number of tasks, preemption introduces more tasks in the network, which results in it completing $\sim$15\% less per request.

Across all experiments, preemption is more frequently invoked when a single task is fully occupying the resources of a device as opposed to multiple tasks sharing the resources. As the scheduler attempts to balance offloaded workloads across all devices, unless there is high resource contention in the network, forcing multiple offloaded tasks to share a device, it is unlikely that the scheduler will invoke preemption on a low-priority task that is allocated two-cores as the scheduler will maximise the cores allocated in a scenario where there is no resource contention.

\subsection{Time Synchronisation}
Time synchronisation posed a significant issue when developing this system, both from the actual synchronisation of the controller and the edge devices and time keeping on the edge devices themselves. We executed the controller on an M1 Macbook which contains a hardware clock. However, the edge devices (Raspberry Pis) do not possess a hardware clock. Instead their clock is implemented via software. During testing of initial versions of the system we did not utilise state updates to inform the controller to remove completed tasks from the network state. Instead, we relied on timing being accurate at all times, which was not the case. This led to edge devices submitting allocation requests for low priority tasks that would fail to receive allocation on their local device as the controller would perceive the time in which they requested resources as being in contention with the completed high priority tasks that spawned them. To mitigate this we deployed an NTP server on the controller that the edge devices were pointed to, additionally, we configured edge devices to explicitly synchronise with the NTP server at system startup and utilised state update messages to inform the controller to remove completed tasks from the network state to provide further safety. The edge devices use the default polling values of NTPd which initial polls every 64 seconds and can adjust up to 1024 seconds depending on the stability of the clock. As our devices are all within the same LAN NTP can guarantee 1 - 2ms of time error between slave clocks (edge devices) and the time master (M1 Macbook). Considering that our time windows at their smallest are within 10s of milliseconds, this level of accuracy remains sufficient for our needs. If the time scale of the system were to be reduced (more powerful devices processing higher workloads with tighter deadlines, with increased throughput rates) sub-millisecond synchronisation may be required. Protocols such as PTP have been shown to achieve accuracy within 100ns. However achieving this level of accuracy requires hardware timestamping which is not feasible as hardware timestamping in WLANs is an area still under exploration. Despite this PTP can still achieve an average accuracy of 7$\mu$s via software timestamping \cite{symm2010ptp} which is more than sufficient to satisfy tighter deadlines .

\subsection{Partitioning}
In this work we utilised horizontal partitioning, another partitioning technique known as vertical partitioning is frequently employed in DNN offloading solutions. It works by partitioning the model itself, processing a portion of the model locally and offloading the remainder with the split in the model typically providing the lowest communication overheads for the model while still benefiting from the reduced processing time of a more powerful remote device. However, in the proposed scenario of offloading in homogeneous edge networks, there is no reduction in processing time to be gained by offloading one portion of the model to another device as they are equal in their computational capabilities.
Instead, in a homogeneous network, offloading is performed to alleviate capacity from overwhelmed edge devices. The only benefit gained by performing vertical partitioning in this scenario is the potential reduction in communication time. However, this
still requires that part of the model is processed locally to
achieve this reduction which may not be feasible when the
local capacity is already overwhelmed.

\subsection{Communication}
The decision to pad communication time-slots to account for network jitter from the initial tests was necessary due to how we estimate network throughput. However, this can lead to poor allocation decisions in the execution of the system if unforeseen interference impacts communication over the air. The scheduler makes strict allocation decisions, when these variances occur they can result in a low priority task arriving on its allocated host late and in turn overrun their allocated processing window. In the event that a task overruns its allotted window the edge device will terminate it, issuing a task violation message to the controller as the edge device needs to provide the occupied resource to the next task that was allocated. 

To examine the impact of throughput variation we evaluated a reduced set of experiments using a more responsive method of throughput estimation using an exponential moving average (EMA) based on actively measured communication times. In all experiments it maintained comparable performance to the static throughput solution which may indicate that when padding is introduced the variation in network throughput is negligible such that that a more responsive approach does not lead to significant changes.

Another issue encountered was the implementation of communication between devices. In this system we utilised REST which resulted in additional delays on both the controller and edge devices as all communication had to be encoded and sent via HTTP. For a distributed system where all communication is controlled, a binary communication format such as GRPC may provide lower communication overheads due to smaller message sizes and simpler data encoding.

\section{Conclusions}
\label{sec:conclusions}
In this paper we applied priority and deadline aware DNN offloading to task pipelines. Overall, preemption leads to better results when performing priority and deadline constrained offloading. Scheduler solutions are able to better integrate with a preemption mechanism and avoid the rash task placement decisions that the workstealing approaches are prone to. One of the main issues with preemption is the impact it has on the completion on sub-tasks within a set when all tasks require completion in order for the set completion to be satisfied. Providing the scheduler with the ability to consider the impact on existing task sets within the network and select the set least likely to complete may mitigate this issue.

However, the main constraint of this system is the hardware it is running on. The Raspberry Pi 2 Model B does not contain the necessary hardware to process DNN tasks in a timely manner. When running the necessary middleware to communicate to the controller and execute several DNN tasks simultaneously across multiple cores, the performance of individual DNN tasks degrades. Even, when using horizontal partitioning to reduce the processing time of a DNN task, it still takes $\sim$14.5s on average to process with a deviation of  $\sim$2.3s. 

Additionally, the slow throughput of 802.11n makes offloading more costly than local processing. Without evaluating this system on another architecture we cannot say for certain but we believe that more powerful devices would result in the computation / communication ratio in skewing further in toward communication but the benefit for preemption would remain as preemption helps mitigate the impact of high volume workloads on a capacity limited network. 
The viability of this approach on these edge devices clearly demonstrates that a more computationally robust architecture would allow for an increase in the volume of work processed by the system and when combined with preemption would allow the system to satisfy even greater workloads.

Considering the tight deadline allocated to high priority tasks (980ms), keeping the search time low for high priority tasks is important. As there are only four edge devices in our network, even under heavily loaded workloads the preemption capable scheduler never exceeds 50ms for allocations that do not require preemption. However, the act of a failed allocation followed by searching for a low priority task to halt and resume high priority allocation results in allocation times as high as $\sim$400ms. As we utilise an M1 Macbook as the controller for our experiments, improving the performance of edge devices would not address this. 

When low-priority tasks are preempted, our system attempts to reallocate them but reallocation is unlikely to occur as there is little time left before their deadline, even with horizontal partitioning allowing for a reduction in processing time, if a preempted task cannot be reallocated the result of the entire frame and the time the network has spent processing it, is rendered redundant. 
A potential avenue for improvement is to eschew reallocation entirely, instead preempting the violating task and it's associated tasks within its set so as to free up the internal task queue and decrease system latency.

In future work we would like to explore the following: using early-exit policies to allow offloaded tasks to complete when a local task requires the resource they occupy, only using preemption as a last resort; preemption selection that considers the benefit to frame completion in the network by selecting tasks from sets that are less likely to complete and finally, more efficient capacity estimation mechanisms as this is a major bottleneck of the scheduler which results in large search times and limits the amount of constraints the scheduler can consider while still remaining performant.

\begin{acks}
This publication has emanated from research conducted with the financial support of Taighde Éireann – Research Ireland under Grant number 18/CRT/6222. For the purpose of Open Access, the author has applied a CC BY public copyright licence to any Author Accepted Manuscript version arising from this submission.
\end{acks}

\bibliographystyle{ACM-Reference-Format}
\bibliography{sample-base}
\end{document}